\documentstyle[aps,epsf,rotate,twocolumn,array]{revtex}
\begin{document} 
\draft 
\flushbottom 
\twocolumn[\hsize\textwidth\columnwidth\hsize\csname 
@twocolumnfalse\endcsname 
 
\title{Diluted quantum antiferromagnets: spin excitations and  
long-range order} 
\author{A. L. Chernyshev$^{1,\dag}$, 
Y. C. Chen$^{2}$, and A. H. Castro Neto$^{3,*}$} 
\address{
$^{1}$Solid State Division, Oak Ridge National Laboratory, Oak Ridge, 
Tennessee 37831\\ 
$^{2}$Department of Physics, University of California, Riverside, 
California 92521\\ 
$^{3}$Department of Physics, Boston University, Boston, MA 02215} 
\date{\today} 
\maketitle 
 
\widetext\leftskip=1.5cm\rightskip=1.5cm\nointerlineskip\small 
\begin{abstract} 
\hspace*{2mm} 
We have studied the static and dynamic magnetic properties of  
two-dimensional (2D) and quasi-two-dimensional, spin-$S$, quantum Heisenberg  
antiferromagnets (QHAF) diluted with spinless vacancies. 
Using spin-wave theory and $T$-matrix approximation we have calculated the 
staggered magnetization, $M(x,T)$, the neutron scattering dynamical structure 
factor, ${\cal S}({\bf k},\omega)$,  the 2D magnetic correlation 
length, $\xi(x,T)$,   
and the N\'eel temperature, $T_N(x)$, for the quasi-2D case. 
We find that  in 2D the  
hydrodynamic description of excitations in terms of spin-waves breaks  
down at the wavelength {\it larger} than $\ell/a \sim  e^{\pi/4x}$, $x$  
being impurity concentration and $a$ the lattice spacing. We 
find the signatures of localization associated with the  
scale $\ell$ and interpret it as the localization length of 
magnons. The spectral function for momenta $a^{-1}\gg k \gg \ell^{-1}$ 
consists of 
two distinct parts: ({\it i}) a damped quasiparticle   
peak at the energy $c_0 k \agt \omega \gg \omega_0$ with abnormal 
damping $\Gamma_k\sim x\, c_0 k$,  
where $\omega_0\sim c_0\ell^{-1}$, $c_0$ is the bare spin-wave
velocity; ({\it ii}) a  
non-Lorentian localization peak at $\omega \sim \omega_0$. For  
$k \alt \ell^{-1}$ these 
two structures merge and the spectrum becomes incoherent. The density of 
states acquires a constant term and 
exhibits an anomalous peak at $\omega \sim \omega_0$ associated with 
the low-energy localized excitations. These anomalies lead to 
a substantial enhancement of the magnetic specific heat, $C_M$, at low 
temperatures.  
Although the dynamical properties are significantly modified we show
that 2D is  
not the lower critical dimension for this problem.  We find that at small $x$ 
the average staggered magnetization at the magnetic site  is 
$M(x,0) \simeq S-\Delta -B\, x$, where $\Delta$ is the 
zero-point spin deviation and $B\simeq 0.21$ is independent of the 
value of $S$; N\'eel temperature
$T_N(x) \simeq (1-A_s\, x)\ T_N(0)$, where $A_s= \pi-2/\pi+ B/(S-\Delta)$ 
is weakly $S$-dependent.  
Our results are in quantitative 
agreement with the recent Monte Carlo simulations 
and experimental data for $S=1/2$, $S=1$, and $S=5/2$.  
In our approach long-range order persists up to a high concentration
of impurities   
$x_c$ which is above 
the classical percolation threshold, $x_p \approx 0.41$. This result suggests 
that long-range order is stable at small $x$ and can be lost only around 
$x\simeq x_p$ where approximations of our approach become invalid.  
\end{abstract} 
] 
\narrowtext 
 
\section{Introduction} 
 
The discovery of superconducting cuprates (HTC) has motivated
an enormous amount of studies in low-dimensional magnetic systems during
the last fifteen years \cite{chn,Manousakis,Kastner}. 
Yet, the superconducting materials now form only a subfield in  
the activity of low-D quantum magnetism (see Ref. \cite{Singh}). 
The hope for an insight into the physics of HTC from the 
study of magnetically related system has attracted much attention to 
the properties of diluted 2D,  QHAF
\cite{Sen,NN,Mahan,KK_imp,Sandvik,Sandvik_bond,Sushkov,Ng,Clarke_th,Oitmaa,kampf,Yasuda,Cheong,Greven,Hucker,Carretta,Carretta1,Clarke,Hammar,Takeda,Kato,anders,anders1}.
One of such systems, La$_{2}$Cu$_{1-x}$Zn(Mg)$_{x}$O$_{4}$ (LCO), a quasi-2D, 
$S=1/2$, QHAF diluted with spinless vacancies, has been a subject of
great interest because of the possibility of new quantum critical points
(QCP) in its phase diagram. 
Earlier experimental data \cite{Cheong}, while demonstrating that  
LCO shows much stronger stability against doping 
in comparison with the mobile hole doped compound 
La$_{2-x}$Sr$_x$CuO$_{4}$, have indicated the existence of a QCP  
at $x=x_c \approx 0.2$, well below the classical percolation  
threshold. This finding was in a sharp contrast with the classical
magnetic systems where dilution leads to the 
breaking of the magnetic bonds and long-range order (LRO) is lost only
at the percolation threshold $x_p$, a characteristic value of the
dilution fraction $x$ at which the last infinite cluster of connected spins
disappears. For a 2D square lattice $x_p \approx 0.41$ \cite{xp}.  
 The existence of such a QCP below the percolation threshold 
was thought to be possible given the large amount of quantum
fluctuations in the ground state of $S=1/2$ system.
However, only recently the systematic experimental analysis  
of the diluted 2D AF has been performed in a wide range of doping  
\cite{Greven,Hucker,Carretta}. Although there are several other 
experimental realizations of a 2D  
QHAF on square lattice with $S=1/2$ \cite{Carretta1,Clarke,Hammar} 
and $S=5/2$ \cite{Takeda,from_Cheong}, 
the CuO-based compounds are among the few which allow a direct probe 
via elastic and inelastic neutron scattering. 
At the same time, quantum Monte Carlo (MC)  
studies have provided highly reliable simulations on large lattices at
low temperatures \cite{Kato,anders}.   
These works indicate that, in fact, no QCP point exists below $x_p$ and 
that at percolation threshold the phase transition is characterized by
classical exponents \cite{anders,anders1}.  
 
Note that the theoretical studies of diluted spin systems 
has attracted much attention  
some 30 years ago in the context of magnetism in diluted magnetic alloys 
\cite{Harris,Izyumov,stinchcombe,Tahir_Kheli}. 
Most of these studies focused 
on the large $S$ (classical) Heisenberg or Ising systems.  
Traditional view of the effect of local  
disorder on the spectrum of an ordered 3D antiferromagnet is that at 
long wavelengths the {\it form} of the spectrum is not modified.  
The only effects are the reduction of hydrodynamic parameters (spin 
stiffness, spin-wave velocity, etc. \cite{HH})   
and a weak damping.  
Conventional arguments for this ability of the long-wavelength 
spectrum to withstand   
perturbations on a local scale often appeal to the Goldstone theorem 
\cite{Shender},  
although its applicability 
to systems without translational invariance requires an 
additional assumption that the microscopic details are virtually
averaged on short distances. As a result the low-energy excitations
are the weakly damped spin-waves which belong to the so called
``infinite cluster'' and they are well defined up to the percolation
threshold \cite{Shender}.
This effective restoration of the translational invariance 
involves the spin-wave propagation on
randomly directed paths with some Euclidean distance $L^\prime$ which
can be converted to a ``true'' distance $L$. Thus, the
wave-vector preserves its meaning at long wavelengths \cite{Ziman}.
In 3D these arguments work very well and are 
assumed to demonstrate a ``general principle''.  

There is growing evidence that in 2D, however, such a logic is not 
always valid. In the 2D case it was found  
by Harris and Kirkpatrick \cite{Harris} and more recently in 
Ref. \cite{Wan}, using a perturbative (linear 
in $x$) approach, that the spin-wave self-energy  
at long wavelengths acquires 
a non-hydrodynamic contribution which explicitly violates the Lorentz 
invariance of the clean system (a feature anticipated by Chakravarty,
{\it et al.}  in Ref. \cite{chn}).  
Recently, similar results have been obtained in RPA studies of 
the diluted 2D Hubbard model \cite{Sen}. 
Some of these studies have concluded that $D=2$ 
is the lower critical dimension for this type of  
disorder\cite{Harris,Wan}, 
implying an instability of the long-range order to an 
infinitesimal doping in the Imry-Ma sense \cite{imry_ma}.  
However, as we mentioned above, MC results show that the order is 
preserved up to $x=x_p$ in contradiction with these conclusions.  
We will show that the conjecture of the instability is an 
artifact of a perturbative 
expansion and is avoided when the divergent series of diagrams is summed. 
However, the resulting modification of the excitation spectrum is 
very unusual and leads to a number of observable anomalies. 
 
Technically, 
our approach is similar to the one by Brenig and Kampf \cite{kampf} 
who have studied the problem of excitation spectrum in diluted 2D QAF 
using spin-wave and $T$-matrix formalism. 
However, while the authors of Ref. \cite{kampf} have noticed 
unusually broad peaks in the spectrum, they {\it assumed} \, a ``normal'' 
3D-type of the spectrum renormalization, that is, the softening of the 
sound velocity and a recovery of the spectrum at long wavelength. 
The derivative of the spin-wave velocity with $x$, obtained 
numerically in Ref. \cite{kampf} using this assumption,  
is rather large $d(c(x)/c_0)/dx \approx -3$ which has supported earlier  
experimental expectations of a QCP at $x < x_p$. 
A recent work by two of us using the non-linear 
$\sigma$-model allied to classical percolation theory \cite{Chen} 
gave similar result.  Another study in Ref. \cite{Man1} used a  
generalizations of the $\sigma$-model with 
parameters modified according to MC data and suggested a 
simple renormalization of spin stiffness $\rho_s(x)$ and spin-wave 
velocity $c(x)$ as the only effect of impurities. 
We will show that these results are not correct 
because of the existence of localized spin excitations which are 
not taken into account in these works.  
 
In this work we study the problem of impurities in 2D QHAF within the linear 
spin-wave theory using the $T$-matrix approach combined with configurational
average over the random positions of impurities. We solve the single impurity 
problem exactly. The spin-wave Green's function is evaluated by 
summing all multiple-scattering diagrams that involve single 
impurity. This approximation gives 
results that go beyond simple linear expansion in $x$, although not 
all higher order  
contributions in $x$ are taken into account. This approach is valid as long as 
single-impurity scattering is the dominant one. We recover the results of 
Refs. \cite{Harris,Wan} at $k \gg \ell^{-1}$, that is, the spin-wave 
spectrum acquires a non-linear logarithmic contribution $\Sigma_{\bf 
k}(\omega)\propto x\, k\, \ln|\omega|$ 
with an abnormal damping  $\Gamma_{\bf k}\propto x \, k$.  
This means that an effective spin-wave velocity $c(x)$ is not well 
defined. However, we show that 
there is no instability of the system towards a disordered phase, 
as conjectured 
previously. The static properties such as staggered magnetization and 
N\'eel temperature do not possess anomalies in contrast with the dynamic 
properties. It is interesting to note that the spin-wave stiffness 
$\rho_s(x)= c(x)^2 \chi_\perp(x)$ is also well defined since the anomalous 
terms in the transverse susceptibility $\chi_\perp(x)$ and in $c(x)$ 
cancel each other. 
 
We show that the diluted 2D AF provides an example of the system
where the arguments for the spectrum to be ``protected'' at long 
wavelengths fail \cite{anderson,pines}.   
We have found that the spectrum of a 2D AF  
at long wavelengths is overdamped at arbitrary concentration   
of spinless impurities. More explicitly, the spectrum ceases to contain a 
quasiparticle peak of any   
kind beyond a certain length-scale. The actual spin excitations 
instead of being described as ballistic may be 
interpreted as diffusive spin modes. 
The reason for that is the influence of scattering centers on the 
long-wavelength excitation   
which is not vanishing in 2D because of the small phase space. This 
leads to the absence   
of the effective self-averaging of the system to a translationally 
invariant medium   
with the renormalized parameters as it would be in 3D.  
Instead, the scattering 
leads to a new length scale  
$\ell/a \sim e^{\pi/4x}$  
beyond which the influence of impurities on the spectrum is dominant.  
We associate this length scale with the localization length of spin 
excitations. 
 
We show that the dynamical structure factor ${\cal S}({\bf k},\omega)$ for 
$a^{-1}\gg k\gg\ell^{-1}$ consists of three
parts (we use units such that $\hbar=k_B=1$): 
({\it i}) a broadened quasiparticle  
peak at the energy $\omega \simeq c_0 k\, (1+2 x \ln(ka)/\pi)$, where 
$c_0= 2 \sqrt{2} S J a$ is the bare spin-wave velocity, $J$ is the 
antiferromagnetic exchange constant, with a width  
given by $\Gamma_k \simeq x \, c_0 k$;  
({\it ii}) a non-Lorentian localization 
peak at  $\omega =\omega_0 \sim c_0\ell^{-1}$, ({\it iii}) a flat
background of states between $\omega=c_0 k$ and $\omega=\omega_0$.  
Thus, besides the lack of the Lorentz invariance, for every ${\bf
k}$-state some weight is spread  from the high energies $\omega \sim
c_0 k$ to the low energies down to $\omega \sim \omega_0$ \cite{remark}. 
For $k \alt \ell^{-1}$ the quasiparticle and localization peaks in 
${\cal S}({\bf k},\omega)$ merge into a broad incoherent peak that
disperses in momentum space. 

The anomalies in 
the dynamical structure factor are reflected in the magnon density of states, 
$N(\omega)$. In a clean 2D AF $N(\omega)\propto\omega$. With the
doping $N(\omega)$ acquires a constant contribution from the localized
states $N(\omega)\propto \omega+ const\cdot x$ at $\omega\gg \omega_0$,
has a peak at $\omega \approx \omega_0$ of the height $\sim 1/x$, and 
vanishes as $N\propto1/(x\ln|\omega|)^2$ for $\omega \ll \omega_0$.  
This behavior of $N(\omega)$ 
is reminiscent of the problem of localization of 
Dirac fermions in 2D $d$-wave  
superconductors \cite{lee} in the case of ``strong'' disorder 
(unitary scatterers). Another interesting similarity between that 
problem and impure 2D AF is that disorder may lead to very different 
physical consequences depending on its ``class''. As it was noted in 
Ref. \cite{Wan} and also in Ref. \cite{Leonid_P} in another context, 
one obtains drastically different results if the spin of impurity 
is equal to the spin of the host material and only bond strengths 
around impurity ($J$) 
are modified. The renormalization of the spectrum in that case does 
not contain any anomalous terms, namely: $\Sigma_{\bf 
k}(\omega)\sim x \, c_0 k $ and $\Gamma_{\bf k} \sim x \, c_0 k^3a^2$. 
According to the terminology of 2D Dirac fermions this problem falls 
into the class of a ``weak'' disorder. 
In the case of spinless impurities the similarity to ``strong'' 
unitary scattering centers is evident since no spin degrees of freedom 
exist at the impurity site. 
  
From the density of states $N(\omega)$ we calculate the magnetic specific  
heat which for a  clean 2D system at low temperatures is  $C_M(0,T) 
\propto T^2$. We predict a strong deviation from this behavior 
due to localized states. We find that specific heat acquires a 
quasi-linear correction $\delta C_M(x,T) =
\beta(x) T/\big(\ln^2|T/\omega_0|+\pi^2/4\big)$  
which is roughly  $ \propto x T$ at  
$T \gg \omega_0$, $\beta(x)\sim 1/x$.
Observation of such a behavior can provide a simple test of our
theory.
We remark that in our approach the contribution of the finite
(decoupled) clusters is not taken into account since the whole system
is considered as a single, ordered, infinite cluster. However, finite
clusters of the size $L$ have a gap in their spectrum of order $J/L$
and thus become important in the low-$T$ region only at $x$ close to
percolation threshold where $L$ can be large. Another source of similar
high-energy corrections is from the resonant states ($\omega_{res}\sim
J$) around impurities
whose energy may go down with doping \cite{CB}.
 At lower temperatures, $T \alt \sqrt{J_\perp J}$,
where $J_{\perp}$ is the inter-plane exchange constant, the
crossover to a 3D behavior should be seen. Thus,  for $x$ not too
close to $x_p$ we expect a large
temperature window where the predicted 
anomalous 2D behavior of the $C_M$ in the infinite cluster is 
dominant and can be observed.
 
We also consider the effects of small inter-plane 
coupling $\tau_{3D}=J_\perp/2J$ and small anisotropy gaps on our 
conclusions for dynamical properties of a strictly 2D isotropic AF we 
discussed above. It is evident that as long as these additional energy 
scales are small in comparison with $J$ there will be an energy range 
$1~\gg~\omega/J~\gg~\sqrt{\tau}$\, ($\tau=\tau_{eff}$ accumulating 
the total effect of the gaps and 3D coupling)  
in which the nonlinearity of the spectrum and an abnormal damping of 
the 2D spin waves should be observable. 
A more delicate question is if the localization part of the spectrum 
and truly overdamped long-wavelength excitations 
can be seen in the presence of gaps or 3D coupling. 
The point is that the disorder induced scale $\omega_0\sim 
J e^{-\pi/4x}$ can be hindered by these additional terms 
which cuts off the log-singularity. 
Therefore a range of concentrations 
$0<x<x^*\sim\ln^{-1}(1/\tau)$ can be found where the long-wavelength 
quasiparticles are still well defined deep in the 3D region of the  
${\bf k}$-space ($ka\ll\sqrt{\tau}$) similar to the 
quasi-1D problem \cite{SK}. 
For the LCO materials $\tau\sim 10^{-4}$ gives $x^*\sim 0.1-0.2$. 
Above the concentration $x^*$ (and at $x<x_p$) localization and 
overdamped peaks should be observable since $\omega_0>\sqrt{\tau}$
and all the low-energy excitations become incoherent. 
Our order of magnitude estimation for the largest value of $\tau$ 
 which can allow such observation (from the condition $x^*<x_p$) is  
$\tau \sim 0.01$. 
Therefore, a rather high impurity concentration
and small enough anisotropies and inter-planar coupling may be required
to observe directly some of the dynamical effects we predict in this
work.  
 
We calculate the static magnetic properties and find a quantitative 
agreement  
with both MC simulations and experimental data. We show that at $T=0$ the 
staggered magnetization (averaged over the magnetic sites 
\cite{remark1}) is given by $M(x,0) \approx S-\Delta-B x$ for  
$x \ll 1$, the factor $\Delta=\sum_k v_{\bf k}^2 \approx 0.2$ stands 
for the contribution of the zero-point fluctuations of the spins, 
$B\simeq 0.21$ is $S$-independent in our approach. 
We find that $T_N(x)/T_N(0) \simeq 1 - A_s\, x $ for $x \ll 1$ where 
$A_s=\pi-2/\pi +B/(S-\Delta)$ is a weak function of $S$.  
This linear expansion result gives $A_{1/2}\simeq 3.2$ and 
$A_{5/2}\simeq 2.6$ which work quite well up to a high value of 
$x\sim 0.25$. It is interesting that the linear expansion results 
point to $x_c(1/2)\simeq 0.31$ and 
$x_c(5/2)\simeq 0.38$,  both below 
$x_p$, which means that  $T_N(x)$ versus $x$ curve should be concave, 
in contrast with the 2D Ising magnets for which  $T_N(x)$ is a 
more traditional convex curve \cite{Cheong}. Such an anomalous
curvature of the ordering temperature has been also 
observed in many different
magnetic systems composed of $f$-electron moments such as U and Ce
\cite{rob}. 
We show that in our approach for larger values of $x$ \, $T_N(x)$ 
indeed bends inward and tends to saturate close to $x_p$. We interpret 
this behavior as due to localization effects which tend to reduce the 
role of quantum fluctuations in the destruction of the long-range order.  

We have calculated the 2D magnetic correlation 
length, $\xi(x,T)$, to describe the paramagnetic phase of the system 
above the N\'{e}el temperature. We used a modified spin-wave theory 
formalism by Takahashi \cite{Takahashi} and calculated  $\xi(x,T)$ 
numerically. Correlation length is suppressed in comparison with the 
pure case and also shows some deviation from the simple $e^{2\pi 
\rho_s(x)/T}$ behavior at larger $x$.  
 
This paper is organized as follows: we describe the model and 
introduce the formalism in Section \ref{formalism};  
in Section \ref{dynamic}, we present the results for the dynamic properties; 
in Section \ref{static} the static properties and long range order is 
discussed; Section \ref{conclusions} contains our conclusions. 
A few Appendices are included with details of the calculations. 
Some of the results presented here were briefly reported in our previous 
paper \cite{Chernyshev}. 
 
 
\section{Formalism} 
\label{formalism} 
 
The systems discussed in this paper are modeled by the site-diluted 
 quantum Heisenberg antiferromagnet:  
\begin{eqnarray} 
H=\sum_{\langle ij\rangle} J_{ij} \, p_{i} \, p_{j} \, {\bf S}_{i}  
\cdot {\bf S}_{j} \, ,   
\label{H1} 
\end{eqnarray} 
where $p_i=1$ ($0$) if ${\bf R}_i$ site is occupied (unoccupied) by the spin 
$S$. We focus on the problem of tetragonal or square lattices with in-plane, 
$J$, and out-of-plane, $J_{\perp}$, nearest-neighbor exchange 
constants, $\langle ij\rangle$ denotes summation over bonds.   
In the systems of interest $J \gg J_{\perp}$ (for instance, in LCO  
$J \approx 1500$ K and $J_{\perp} \sim 10^{-4} J$). 
 
\subsubsection{Spin-wave approximation.} 
We begin with the Hamiltonian (\ref{H1}) which is split into the pure 
host and impurity part 
\begin{eqnarray} 
\label{H1a} 
{\cal H}={\cal H}_0+{\cal H}_{imp}= 
\sum_{\langle ij\rangle} J_{ij}\, {\bf S}_{i} \cdot {\bf S}_{j} 
-\sum_{l,\delta} J_{l,\delta}\, {\bf S}_{l} \cdot {\bf S}_{l+\delta}\ , 
\end{eqnarray} 
where $l$ runs over the impurity sites and $\delta$ is a 
nearest-neighbor unity vector.  
Then, in the linear spin-wave approximation, 
\begin{eqnarray} 
\label{S_to_a} 
&& S_{i}^{z}= S-a_{i}^{\dagger}a_{i}, \  
S^+_{i}\simeq\sqrt{2S}a_{i}, \ S_{i}^{-}\simeq 
\sqrt{2S}a_{i}^{\dagger} \ , 
\nonumber \\ 
&& S_{j}^{z}= -S+b_{j}^{\dagger}b_{j}, \ 
S^+_{j}\simeq\sqrt{2S}b_{j}^{\dagger}, \ S_{j}^{-}\simeq \sqrt{2S} 
b_{j}\ , 
\end{eqnarray} 
for the spins in $A$ ($i$) and $B$ ($j$) sublattices 
quadratic part of the pure host Hamiltonian ${\cal H}_0$ for the 
tetragonal lattice is given by: 
\begin{eqnarray} 
\label{H_0} 
&&{\cal H}_0 = 4SJ\sum_{\bf k} 
\bigg[\hat{\gamma}_0 (a^{\dag}_{\bf k}a_{\bf k}+b^{\dag}_{\bf k} 
b_{\bf k})\nonumber \\ 
&&\phantom{{\cal H}_0 = 4SJ\sum_{\bf k}\big[} 
+\hat{\gamma}_{\bf k} 
(a^{\dag}_{\bf k}b^{\dag}_{\bf -k}+b_{-{\bf k}}a_{\bf k})\bigg]\ , 
\end{eqnarray} 
where we use that in-plane and out-of-plane coordination numbers 
are $z=4$ and $z_\perp=2$, respectively, and define 
\begin{eqnarray} 
\label{hat_gamma} 
\hat{\gamma}_{\bf k}=\gamma_{\bf k}+\tau\gamma_{\bf k}^{\perp}\ , 
\end{eqnarray} 
with $\tau=J_\perp/2J$, $\gamma _{\bf k}=\left(\cos k_x+\cos 
k_y\right)/2$, and $\gamma_{\bf k}^{\perp}=\cos k_z$.  
From now on the in-plane and out-of-plane momenta are in units of the
correspondent inverse lattice constants.
Impurity part of the Hamiltonian (\ref{H1a}) on the tetragonal 
lattice is:  
\begin{eqnarray} 
\label{H_imp_r} 
&&{\cal H}_{imp}^A=-S\sum_{l\in A,\delta}J_{l,\delta} 
\bigg[a^{\dag}_l a_l+b^{\dag}_{l+\delta} 
b_{l+\delta}+a^{\dag}_l b^{\dag}_{l+\delta}+a_l b_{l+\delta}\bigg]  
\ ,\nonumber \\ 
&&{\cal H}_{imp}^B={\cal H}_{imp}^A(a\leftrightarrow b)\ , 
\end{eqnarray} 
with $J_{l,\delta}=J$ ($J_\perp$) for $\delta=e_x, e_y$ ($e_z$). 
After Fourier transformation it is more convenient to write impurity 
Hamiltonian in the $2\times 2$ matrix notations: 
\begin{eqnarray} 
\label{H_imp_k} 
{\cal H}_{imp}=-4SJ\sum_{l,{\bf k},{\bf k}^\prime} 
e^{i({\bf k}-{\bf k}^\prime){\bf R}_l} \hat{A}_{\bf k}^{\dag} 
\hat{V}^l_{{\bf k},{\bf k}^\prime} \hat{A}_{{\bf k}^\prime} \ , 
\end{eqnarray} 
where 
\begin{equation} 
\hat{A}_{\bf k} = 
\left[ 
\begin{array}{c} 
a_{\bf k} \\  
b_{-{\bf k}}^{\dag} 
\end{array} 
\right]\, ,\ \  \hat{A}_{\bf k}^{\dag} = \left[ 
\begin{array}{cc} 
a_{\bf k}^{\dag}, &  
b_{-{\bf k}} 
\end{array} 
\right]\, , 
\end{equation} 
with scattering potentials for $l$ in the sublattice $A$: 
\begin{equation} 
\hat{V}^A_{{\bf k},{\bf k}^\prime}=\left(  
\begin{array}{cc} 
\hat{\gamma}_0 & \hat{\gamma}_{{\bf k}^\prime}\\ 
\hat{\gamma}_{\bf k} & 
\hat{\gamma}_{{\bf k}-{\bf k}^\prime} 
\end{array} 
\right) , 
\label{8} 
\end{equation} 
and for $l$ in the sublattice $B$: 
\begin{equation} 
\hat{V}_{{\bf k},{\bf k}^\prime}^{B}=\left(  
\begin{array}{cc} 
\hat{\gamma}_{{\bf k}-{\bf k}^\prime} &  
\hat{\gamma_{\bf k}} \\  
\hat{\gamma}_{{\bf k}^\prime} & \hat{\gamma}_0 
\end{array} 
\right) \, .  \label{9} 
\end{equation} 
The pure host Hamiltonian (\ref{H_0}) is diagonalized using Bogolyubov 
transformation: 
\begin{eqnarray} 
\label{BT} 
&&a_{\bf k}=u_{\bf k} 
\alpha_{\bf k}+v_{\bf k}\beta^{\dag}_{-{\bf k}}\ ,\nonumber\\ 
&&b^{\dag}_{\bf k}=u_{\bf k} 
\beta^{\dag}_{\bf k}+v_{\bf k}\alpha_{-{\bf k}} \ , 
\end{eqnarray} 
with  
\begin{eqnarray} 
\label{BT1} 
&&u_{\bf k}^2-v_{\bf k}^2=1\ ,\ \  
2u_{\bf k}v_{\bf k}= -\hat{\gamma}_{\bf k}/\omega_{\bf k}\ ,  
\\  
&&u_{\bf k}=\sqrt{\frac{\hat{\gamma}_0+\omega _{\bf k}}{2\omega _{\bf 
k}}}\ , \ \ 
v_{\bf k}=-\mbox{sgn}\,\hat{\gamma}_{\bf k} 
\sqrt{\frac{\hat{\gamma}_0-\omega _{\bf k}}{2\omega _{\bf k}}}\ , \nonumber 
\end{eqnarray} 
where bare spin-wave frequency is 
\begin{eqnarray} 
\label{wk} 
\omega_{\bf k}=\sqrt{\hat{\gamma}_0^{2}-\hat{\gamma}_{\bf k}^{2}} \ . 
\end{eqnarray} 
The problem can be reduced to the problem in 2D square lattice 
by letting $\tau\rightarrow 0$ in 
Eqs. (\ref{H_0})-(\ref{wk}).   
In what follows all energies are expressed in the units of 
$\Omega_0=4SJ$.  
 
After the Bogolyubov transformation the Hamiltonian 
Eqs. (\ref{H_0}),(\ref{H_imp_k}) is given by (in the units of $\Omega_0$): 
\begin{eqnarray} 
\label{H_BT_0} 
&&{\cal H}_0=\sum_{\bf k} \omega_{\bf k} \left(\alpha^{\dag}_{\bf k} 
\alpha_{\bf k}+\beta^{\dag}_{\bf k} 
\beta_{\bf k}\right)\ ,\\ 
\label{H_BT} 
&&{\cal H}_{imp}=-\sum_{l,{\bf k},{\bf k}^\prime} 
e^{i({\bf k}-{\bf k}^\prime){\bf R}_l} {\hat{\cal A}}_{\bf k}^{\dag} 
{\hat{\cal V}}^l_{{\bf k},{\bf k}^\prime} {\hat{\cal A}}_{{\bf k}^\prime} \ , 
\end{eqnarray} 
where two-component vectors are: 
\begin{equation} 
{\hat{\cal A}}_{\bf k} 
= 
\left[ 
\begin{array}{c} 
\alpha_{\bf k} \\  
\beta_{-{\bf k}}^{\dag} 
\end{array} 
\right]\, ,\ \   
{\hat{\cal A}}_{\bf k}^{\dag} 
= \left[ 
\begin{array}{cc} 
\alpha_{\bf k}^{\dag}, &  
\beta_{-{\bf k}} 
\end{array} 
\right]\, , 
\end{equation} 
and $2\times 2$ scattering potential matrices 
${\hat{\cal V}}_{{\bf k},{\bf k}^\prime}$ are obtained from 
Eqs. (\ref{8}),(\ref{9}) using Eq. (\ref{BT}). 
For the sake of the further use of the $T$-matrix formalism 
it is convenient to decompose scattering potentials into the orthogonal 
components according to the symmetry with respect to the scattering 
site. The symmetry of the tetragonal lattice is $D_{4h}$, which is a 
group of order 16 and has 10 irreducible representations.  
Since the impurity potentials Eqs. (\ref{8},\ref{9}) 
connect only nearest-neighbor sites, only five components of the 
scattering potentials in the irreducible representations of $D_{4h}$ are 
nonzero. They correspond to the irreducible representations $A_{1g}$, 
$B_{1g} $, $B_{2g}$, and $E_{u}$. These nonzero 
components are the $s$-wave, in-plane $p_x$-, $p_y$-, and $d$-waves,  
and out-of-plane $p_z$-wave (for details see Appendix \ref{app_A}). 
 
Thus, the scattering potential for the impurity in the sublattice $A$: 
\begin{eqnarray} 
\label{V_A} 
{\hat{\cal V}}^A_{{\bf k},{\bf k}^\prime}=\sum_\mu  
{\hat{\cal V}}^{A,\mu}_{{\bf k},{\bf k}^\prime}\ , 
\end{eqnarray} 
where scattering channels are $\mu=s, p_x, p_y, d, p_z$.  
In each channel the scattering potentials can be written as a  
direct product of the column and row vectors. The $s$-wave part 
\begin{eqnarray} 
\label{V_s} 
&&{\hat{\cal V}}^{A,s}_{{\bf k},{\bf k}^\prime}=|s_{\bf k}\rangle 
\otimes\langle s_{{\bf k}^\prime}|+\tau |s^\perp_{\bf k}\rangle 
\otimes\langle s^\perp_{{\bf k}^\prime}|\ ,\nonumber\\ 
&&\phantom{{\hat{\cal V}}^{A,s}_{{\bf k},{\bf k}^\prime}=} 
\mbox{where} \ \ \langle s_{\bf k}|=  
\big[u_{\bf k}+v_{\bf k}\gamma_{\bf k},\ \ v_{\bf k}+u_{\bf
k}\gamma_{\bf k}\big],  
\\ 
&&\phantom{{\hat{\cal V}}^{A,s}_{{\bf k},{\bf k}^\prime}=} 
 \mbox{and} \ \ \ \ \ \langle s^\perp_{\bf k}|=\big[u_{\bf k}+v_{\bf k} 
\gamma^\perp_{\bf k},\ \ v_{\bf k}+u_{\bf k}\gamma^\perp_{\bf k}\big], 
\nonumber 
\end{eqnarray} 
the in-plane $p$-wave part 
\begin{eqnarray} 
\label{V_p} 
&&{\hat{\cal V}}^{A,p_{x(y)}}_{{\bf k},{\bf k}^\prime}=|p^{x(y)}_{\bf 
k}\rangle\otimes \langle p^{x(y)}_{{\bf k}^\prime}|\ ,\nonumber\\ 
&&\phantom{{\hat{\cal V}}^{A,p_{x(y)}}_{{\bf k},{\bf k}^\prime}=} 
\mbox{where} \ \ \langle p^{x(y)}_{\bf k}|= \sin k_{x(y)} 
\big[v_{\bf k},\  \  u_{\bf k}\big]/\sqrt{2},  
\end{eqnarray} 
the $d$-wave part 
\begin{eqnarray} 
\label{V_d} 
&&{\hat{\cal V}}^{A,d}_{{\bf k},{\bf k}^\prime}=|d_{\bf 
k}\rangle\otimes \langle d_{{\bf k}^\prime}|\ ,\nonumber\\ 
&&\phantom{{\hat{\cal V}}^{A,d}_{{\bf k},{\bf k}^\prime}=} 
\mbox{where} \ \ \langle d_{\bf k}|= \gamma^-_{\bf k} 
\big[v_{\bf k},\  \  u_{\bf k}\big] , \\ 
&&\phantom{{\hat{\cal V}}^{A,d}_{{\bf k},{\bf k}^\prime}=} 
\mbox{with} \ \ \ \ \  \gamma^-_{\bf k}=(\cos k_x - \cos k_y)/2,\nonumber 
\end{eqnarray} 
and the out-of-plane $p_z$-wave contribution  
\begin{eqnarray} 
\label{V_pz} 
&&{\hat{\cal V}}^{A,p_z}_{{\bf k},{\bf k}^\prime}=\tau|p^z_{\bf 
k}\rangle \otimes\langle p^z_{{\bf k}^\prime}|\ ,\nonumber\\ 
&&\phantom{{\hat{\cal V}}^{A,p_z}_{{\bf k},{\bf k}^\prime}=} 
\mbox{where} \ \ \langle p^z_{\bf k}|= \sin k_z 
\big[v_{\bf k},\  \  u_{\bf k}\big]/\sqrt{2} .  
\end{eqnarray} 
For the impurity in $B$ sublattice 
${\hat{\cal V}}^{B,\mu}_{{\bf k},{\bf k}^\prime}\equiv 
{\hat{\cal V}}^{A,\mu}_{{\bf k},{\bf k}^\prime}(u\leftrightarrow v)$. 
 
In what follows we consider the 2D ($\tau=0$) or quasi-2D  ($\tau\ll 
1$) limit of the problem. It can be shown that the contribution  
of the out-of-plane  terms in $s$-wave and 
$p_z$-wave  scattering potentials 
which  explicitly depend on $\tau$, as well as  the one of 
the majority of the $\tau$-dependent terms originating from the 
quasi-2D form of $u_{\bf k}$, $v_{\bf k}$, and $\omega_{\bf k}$ 
(\ref{BT1}), (\ref{wk}) is negligible in the quasi-2D case 
($\sim {\cal O}(\tau)$, see Appendix \ref{app_B}). 
It allows one to simplify the scattering problem further by 
neglecting the $s^\perp$ and $p_z$ components in the above 
equations. Moreover, the solution for the 2D problem can be applied 
directly to the quasi-2D case since the formal expressions are 
identical in both cases. The only important difference concerns the 
logarithmically divergent terms which in quasi-2D system acquire a 
low-energy cut-off provided by the implicit dependence of the scattering 
potentials (\ref{V_p})-(\ref{V_sa}) on $\tau$ through $\omega_{\bf 
k}$. This simply means that for the quasi-2D case  
in the limit $\tau\ll 1$ one can restrict oneself by considering 
purely 2D scattering including the 3-dimensionality only on the level of the  
spin-wave dispersion in certain terms. Thus, in the following we use 
\begin{eqnarray} 
\label{V_sa} 
&&{\hat{\cal V}}^{A,s}_{{\bf k},{\bf k}^\prime}=|s_{\bf k}\rangle 
\otimes\langle s_{{\bf k}^\prime}|\ ,\nonumber\\ 
&&\phantom{{\hat{\cal V}}^{A,s}_{{\bf k},{\bf k}^\prime}=} 
\mbox{with} \ \ \langle s_{\bf k}|=\omega_{\bf k} 
\big[u_{\bf k},\ \ -v_{\bf k}\big]\ . 
\end{eqnarray} 
 The rest of this section is  
devoted to the 2D limit of the problem and, unless specified 
otherwise, we use 
\begin{eqnarray} 
\label{2D_gamma} 
&&\hat{\gamma}_{\bf k}=\gamma_{\bf k}=\left(\cos k_x+\cos 
k_y\right)/2\ , \nonumber\\  
&&\hat{\gamma}_0=1\ , \ \ 
\omega _{{\bf k}}=\sqrt{1-\gamma_{\bf k}^{2}} \ . 
\end{eqnarray} 
 
\subsubsection{$T$-matrix. Single-impurity scattering.}  

We are interested in the Green's function of the Hamiltonian 
(\ref{H_BT_0}) modified by random impurity potentials 
(\ref{H_BT}). The Green's function is a $2\times 2$ matrix defined in 
a standard way: 
\begin{eqnarray} 
\label{G} 
&&G_{\bf k}^{11}(t)=-i\big\langle T[\alpha_{\bf k}(t)\alpha_{\bf 
k}^{\dag}(0)]\big\rangle , \nonumber\\  
&&G_{\bf k}^{12}(t)=-i\big\langle T[\alpha_{\bf k}(t)\beta_{-{\bf 
k}}(0)]\big\rangle , \nonumber\\  
&&G_{\bf k}^{21}(t)=-i\big\langle T[\beta_{-{\bf k}}^{\dag}(t) 
\alpha_{\bf k}^{\dag}(0)]\big\rangle , \\  
&&G_{\bf k}^{22}(t)=-i\big\langle T[\beta_{\bf k}^{\dag}(t) 
\beta_{\bf k}(0)]\big\rangle  ,\nonumber 
\end{eqnarray} 
where brackets also imply a configurational average over the impurity sites. 
 
The $T$-matrix equation for the Hamiltonian (\ref{H_BT}) is given by  
\begin{eqnarray} 
\label{T} 
\hat{T}^{l,\mu}_{{\bf k},{\bf k}^\prime}(\omega)  
=-{\hat{\cal V}}^{l,\mu}_{{\bf k},{\bf k}^\prime}  
-\sum_{\bf q}{\hat{\cal V}}^{l,\mu}_{{\bf k},{\bf q}} \hat{G}^0_{\bf q} 
(\omega) \hat{T}^{l,\mu}_{{\bf q},{\bf k}^\prime}(\omega)\ , 
\end{eqnarray} 
with $l=A(B)$, partial waves are restricted to in-plane  
$\mu = s,p_\sigma,d$ harmonics according to the above discussion,  
and $\hat{G}^0_{\bf q}(\omega)$ is the $2\times 2$  
bare Green's function:  
\begin{eqnarray} 
\label{G_0} 
&&G^{0,11}_{\bf q}(\omega)=G^{0,22}_{\bf q}(-\omega)= 
\frac{1}{\omega-\omega_{\bf q}+i0}\ , \\ 
&&G^{0,12}_{\bf q}(\omega)=G^{0,21}_{\bf q}(\omega)=0\ . 
\nonumber 
\end{eqnarray} 
The diagrammatic equivalent of the Eq. (\ref{T}) is shown in 
Fig. \ref{fig_1}(a). The $T$-matrix equations (\ref{T}) with potentials 
(\ref{V_p})-(\ref{V_sa}) can be readily solved: 
\begin{eqnarray} 
\label{T_1} 
&&\hat{T}^{A,\mu}_{{\bf k},{\bf k}^\prime}(\omega)  
={\hat{\cal V}}^{A,\mu}_{{\bf k},{\bf k}^\prime}  
\Gamma_\mu(\omega), \\ 
&&\hat{T}^{B,\mu}_{{\bf k},{\bf k}^\prime}(\omega)  
={\hat{\cal V}}^{B,\mu}_{{\bf k},{\bf k}^\prime}  
\Gamma_\mu(-\omega), 
\nonumber 
\end{eqnarray} 
where the frequency dependent parts are given by 
\begin{eqnarray} 
\label{gammas} 
&&\Gamma_{s}(\omega)=\frac{1}{\omega}+\frac{(1+\omega)\rho(\omega)} 
{1-\omega(1+\omega)\rho(\omega)}\ ,\nonumber\\ 
&&\Gamma_{p}(\omega)=-\frac{2}{1+\omega+(1-\omega)(\omega^2\rho(\omega) 
-\rho_d(\omega))}\ ,\\ 
&&\Gamma_{d}(\omega)=-\frac{1}{1+(1-\omega)\rho_d(\omega)} 
\ ,\nonumber 
\end{eqnarray} 
with  
\begin{eqnarray} 
\label{rhos} 
\rho(\omega)=\sum_{\bf p}\frac{1}{\omega^2-\omega_{\bf p}^2}\ , \ \ 
\rho_d(\omega) 
=\sum_{\bf p}\frac{(\gamma^-_{\bf p})^2}{\omega^2-\omega_{\bf p}^2}\ .   
\end{eqnarray} 
We note here that the second term in $s$-wave $\Gamma_s(\omega)$ is 
proportional to $\rho(\omega)$ at $\omega \ll 1$, where the latter  
appears 
naturally from the summation in Fig. \ref{fig_1}(a) as a result of 
combination of $G^{0,11}_{\bf q}(\omega)$ and $G^{0,22}_{\bf 
q}(\omega)$ in the internal part of the diagrams.  
When the summation over ${\bf p}$ in Eq. (\ref{rhos}) 
is restricted to 2D $\rho(\omega)$ is a logarithmic function at low 
energies. In the following we show that this contribution to the 
$s$-wave scattering is solely responsible for all the anomalies in the  
spectrum of a 2D AF. Interestingly, similar logarithmic term in the 
self-energy of the 2D Dirac fermions in the problem of disorder in 
$d$-wave superconductors requires a summation of the specific subset 
of diagrams \cite{Yashenkin}.  
In our case, while one needs to sum infinite series of diagrams, no 
special selection or inclusion of the multiple-impurity scattering 
processes is necessary. Since the single-particle density of states 
and sensitivity of the results to the type of disorder in both 
problems are similar, establishing of the detailed correspondence between 
these two problems is an important question.  
Integrals in Eq. (\ref{rhos}) can be taken analytically and,   
in the case of 2D, are expressed through the complete elliptic 
integrals \cite{kampf} (see Appendix \ref{app_C}). 
 
The first term in the $s$-wave scattering Eq. (\ref{gammas}) 
represents a singular zero-frequency mode which is independent of the 
dimension of the problem and originates from the oscillations of the  
fictitious spin degrees of freedom at the impurity site which are 
decoupled from the AF matrix. Roughly speaking, when the spins are 
quantized as in Eq. (\ref{S_to_a}) and $S^\prime$ at the impurity site 
is set to zero there is still $a^\dag_0 a_0$ left from $S^z_0$. Thus, in 
the spin-wave approximation, it gives rise to the $s$-wave zero-frequency mode. 
This problem has been noticed since the earliest works on the diluted magnets  
which have used the spin-wave theory \cite{Izyumov} and also more 
recently in the context of diluted AF 
\cite{Mahan,kampf,Wan,Leonid_P,Bulut} (for an extensive discussion 
see Ref. \cite{Wan}). 
Since these states are unphysical and are unrelated to the low-energy 
physics of the AF they have to be projected out. One of the projection 
schemes involves a non-Hermitian potential which was designed to 
preserve the simplest factorized form of the $s$-wave scattering 
potential \cite{Wan}. We use another, physically more transparent scheme  
which introduces a fictitious magnetic fields at the impurity 
sites (similar to Refs. \cite{kampf,Leonid_P}):  
\begin{eqnarray} 
\label{D_H} 
&&\Delta{\cal H}=H_z\sum_l a^{\dag}_l a_l \\ 
&&\phantom{\Delta{\cal H}}\Rightarrow H_z\sum_{l,{\bf k},{\bf k}^\prime} 
e^{i({\bf k}-{\bf k}^\prime){\bf R}_l} {\hat{\cal A}}_{\bf k}^{\dag} 
\Delta{\hat{\cal V}}^{l,s}_{{\bf k},{\bf k}^\prime}  
{\hat{\cal A}}_{{\bf k}^\prime} \ , \nonumber 
\end{eqnarray}  
where corrections to the $s$-wave scattering potential are: 
\begin{eqnarray} 
\label{D_V_s} 
&&\Delta{\hat{\cal V}}^{A,s}_{{\bf k},{\bf k}^\prime}= 
|\Delta s_{\bf k}\rangle 
\otimes\langle \Delta s_{{\bf k}^\prime}|\ ,\nonumber\\ 
&&\phantom{\Delta{\hat{\cal V}}^{A,s}_{{\bf k},{\bf k}^\prime}=} 
\mbox{with} \ \ \langle \Delta s_{\bf k}|=  
[u_{\bf k},\ \ v_{\bf k}] \ , 
\end{eqnarray}  
$\Delta{\hat{\cal V}}^{B,s}_{{\bf k},{\bf k}^\prime}= 
\Delta{\hat{\cal V}}^{A,s}_{{\bf k},{\bf k}^\prime}\{u\leftrightarrow 
v\}$.  
Evidently, $p$ and $d$ waves are not affected by the projection. 
Within our approach after some algebra in the 
limit $H_z\rightarrow\infty$ one obtains a modified $T$-matrix solution 
(for the case of an arbitrary $H_z$ see Appendix \ref{app_D}):  
\begin{eqnarray} 
\label{T_2} 
&&\hat{T}^{A,s}_{{\bf k},{\bf k}^\prime}(\omega)  
={\hat{\cal V}}^{A,s}_{{\bf k},{\bf k}^\prime}  
\Gamma_s(\omega)+\Delta\hat{T}^{A,s}_{{\bf k},{\bf k}^\prime}(\omega), \\ 
&&\hat{T}^{B,s}_{{\bf k},{\bf k}^\prime}(\omega)  
={\hat{\cal V}}^{B,s}_{{\bf k},{\bf k}^\prime}  
\Gamma_s(-\omega)+\Delta\hat{T}^{B,s}_{{\bf k},{\bf k}^\prime}(\omega), 
\nonumber 
\end{eqnarray} 
where ${\hat{\cal V}}^{l,s}_{{\bf k},{\bf k}^\prime}$  
is given, as before, by Eq. (\ref{V_s}) and  
the frequency dependent part is now free from the zero-frequency pole 
\begin{eqnarray} 
\label{gamma_s} 
\Gamma_{s}(\omega)=\frac{(1+\omega)\rho(\omega)} 
{1-\omega(1+\omega)\rho(\omega)}\ . 
\end{eqnarray} 
Comparing this expression with Eq. (\ref{gammas}) one may note that 
the ``physical'' term is left unchanged after the projection. 
Additional terms in the solution (\ref{T_2}) are also regular: 
\begin{eqnarray} 
\label{D_T} 
&&\Delta\hat{T}^{A,s}_{{\bf k},{\bf k}^\prime}(\omega)=-\omega 
|\Delta s_{\bf k}\rangle 
\otimes\langle \Delta s_{{\bf k}^\prime}|\\ 
&&\phantom{\Delta\hat{T}^{A,s}_{{\bf k},{\bf k}^\prime}(\omega)=} 
+|s_{\bf k}\rangle 
\otimes\langle \Delta s_{{\bf k}^\prime}|+ 
|\Delta s_{\bf k}\rangle 
\otimes\langle  s_{{\bf k}^\prime}|\ , \nonumber
\end{eqnarray} 
with $|s_{\bf k}\rangle$ from Eq. (\ref{V_sa}) and  
$|\Delta s_{\bf k}\rangle$ from Eq. (\ref{D_V_s}), 
$\hat{T}^{B,s}_{{\bf k},{\bf k}^\prime}(\omega)= 
\hat{T}^{A,s}_{{\bf k},{\bf k}^\prime} 
(-\omega)\{u\leftrightarrow v\}$ as before. 
Thus, the projection (\ref{D_H}) allows one to remove the unphysical 
divergency at $\omega=0$ which would otherwise affect the true  
low-energy physics of the problem. 
 
\subsubsection{Green's function.} 
The averaging over random distribution of impurities  
readily transforms $T$-matrix into the spin-wave 
self-energies:  
\begin{eqnarray} 
\label{Sigma_1} 
\hat{\Sigma}_{\bf k}(\omega)= 
\sum_\mu\hat{\Sigma}_{\mu,{\bf k}}(\omega) \ , 
\end{eqnarray} 
with $\mu$-wave contributions 
\begin{eqnarray} 
\label{Sigma_2} 
\hat{\Sigma}_{\mu,{\bf k}}(\omega)=x 
\delta_{{\bf k}-{\bf k}^\prime}\left[\hat{T}^{A,\mu}_{{\bf k},{\bf 
k}^\prime}(\omega)+\hat{T}^{B,\mu}_{{\bf k},{\bf k}^\prime}(\omega) 
\right]\ . 
\end{eqnarray} 
For the 2D case the contribution of the partial waves to 
the self-energies are: 
\begin{eqnarray} 
\label{sigma_s} 
&&\hat{\Sigma}_{s,{\bf k}}(\omega)=x\omega_{\bf k}\biggl[ 
\left( 
\begin{array}{cc} 
1 & \gamma_{\bf k}\\ \gamma_{\bf k} & 1  
\end{array}\right) 
\frac{\Gamma_{s}(\omega)+\Gamma_{s}(-\omega)}{2}\\ 
&&\phantom{\hat{\Sigma}_{s,{\bf k}}(\omega)=-x\omega_{\bf k}}  
+ \left( 
\begin{array}{cc} 
\omega_{\bf k} & 0 \\ 0 & -\omega_{\bf k} 
\end{array}\right) 
\frac{\Gamma_{s}(\omega)-\Gamma_{s}(-\omega)}{2}\nonumber\\ 
&&\phantom{\hat{\Sigma}_{s,{\bf k}}(\omega)=-x\omega_{\bf k}}  
-\left( 
\begin{array}{cc} 
2 & 0 \\ 0 & 2 
\end{array}\right)\biggr]-x\omega 
\left( 
\begin{array}{cc} 
1 & 0 \\ 0 & -1 
\end{array}\right)\ , \nonumber\\ 
\label{sigma_p} 
&&\hat{\Sigma}_{p,{\bf k}}(\omega)=x\omega_{\bf 
k}\left[1-\left(\frac{\gamma_{\bf k}^-}{\omega_{\bf 
k}}\right)^2\right]\\ 
&&\phantom{\hat{\Sigma}_{p,{\bf k}}(\omega)=-x\omega_{\bf k}}  
\times\biggl[\left( 
\begin{array}{cc} 
1 & -\gamma_{\bf k}\\ -\gamma_{\bf k} & 1  
\end{array}\right) 
\frac{\Gamma_{p}(\omega)+\Gamma_{p}(-\omega)}{2}\nonumber\\ 
&&\phantom{\hat{\Sigma}_{p,{\bf k}}(\omega)=-x\omega_{\bf k}\times\biggl[}  
+ \left( 
\begin{array}{cc} 
-\omega_{\bf k} & 0 \\ 0 & \omega_{\bf k} 
\end{array}\right) 
\frac{\Gamma_{p}(\omega)-\Gamma_{p}(-\omega)}{2} 
\biggr]\ ,\nonumber\\ 
\label{sigma_d} 
&&\hat{\Sigma}_{d,{\bf k}}(\omega)=x\omega_{\bf 
k}\left(\frac{\gamma_{\bf k}^-}{\omega_{\bf k}}\right)^2 
\biggl[ 
\left( 
\begin{array}{cc} 
1 & -\gamma_{\bf k}\\ -\gamma_{\bf k} & 1  
\end{array}\right) 
\frac{\Gamma_{d}(\omega)+\Gamma_{d}(-\omega)}{2}\nonumber\\ 
&&\phantom{\hat{\Sigma}_{d,{\bf k}}(\omega)=-x\omega_{\bf k}}  
+ \left( 
\begin{array}{cc} 
-\omega_{\bf k} & 0 \\ 0 & \omega_{\bf k} 
\end{array}\right) 
\frac{\Gamma_{d}(\omega)-\Gamma_{d}(-\omega)}{2} 
\biggr]\ . 
\end{eqnarray} 
It is interesting to observe that  
``on shell'' (at $\omega=\omega_{\bf k}$)  
``projected'' $\hat{T}^{s}_{{\bf k},{\bf k}^\prime}(\omega)$
from Eqs. (\ref{T_2})-(\ref{D_T}) 
and ``non-projected'' expressions Eq. (\ref{T_1}), (\ref{gammas}) 
yield identical 
$\Sigma_{s,{\bf k}}(\omega_{\bf k})$.  
\noindent
\begin{figure} 
\unitlength 1cm 
\epsfxsize=8.cm 
\begin{picture}(8,6.5) 
\put(0,0.5){\epsffile{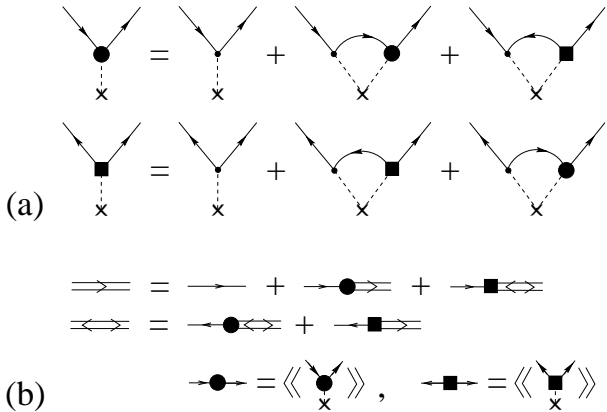}} 
\end{picture} 
\caption{(a) $T$-matrix single-impurity  
scattering series, 
(b) Dyson-Belyaev diagram series for the diagonal,  
$G^{11}$, and 
off-diagonal,  $G^{12}$, Green's functions. Self-energies  
$\Sigma^{11}$ (circle) and $\Sigma^{12}$ 
(square) are the configurational averages of $T^{11}$ 
and $T^{12}$ components of the $T$-matrix, respectively.} 
\label{fig_1} 
\end{figure} 
\noindent 
 
Summation of the Dyson-Belyaev  
diagrammatic series for the Green's functions shown in 
Fig. \ref{fig_1}(b) with self-energies defined in 
Eqs. (\ref{Sigma_1})-(\ref{sigma_d}) gives 
\begin{eqnarray} 
\label{G_full} 
&&\hat{G}_{\bf k}(\omega)=\left(  
\begin{array}{cc} 
-\omega -\omega _{\bf k}-\Sigma_{\bf k}^{22}(\omega) & 
\Sigma_{\bf k}^{12}(\omega) \\  
\Sigma_{\bf k}^{21}(\omega) & \omega -\omega _{\bf k}-\Sigma_{\bf 
k}^{11}(\omega)  
\end{array} 
\right) \\ 
&&\times  \frac{1}{\left(\omega -\omega _{\bf k}- 
\Sigma_{\bf k}^{11}(\omega)\right)\left( -\omega -\omega _{\bf k}- 
\Sigma_{\bf k}^{22}(\omega)\right)-  
\left(\Sigma_{\bf k}^{12}(\omega)\right)^2 }  \nonumber\ , 
\end{eqnarray} 
where $\Sigma_{\bf k}^{22}(\omega)=\Sigma_{\bf k}^{11}(-\omega)$.  
A detailed consideration of the properties of spectral functions  
\begin{eqnarray} 
\label{A_ij} 
A^{ij}_{\bf k}(\omega)=-\frac{1}{\pi} 
\mbox{Im}{\hat{G}}^{ij}_{\bf k}(\omega)\ , 
\end{eqnarray} 
will be given in the next section. 
 
We investigate the neutron scattering dynamical structure  
factor, ${\cal S}({\bf k},\omega)$: 
\begin{eqnarray} 
\label{S_1} 
{\cal S}^{\alpha\beta}({\bf k},\omega)= \int_{-\infty}^\infty dt\,  
e^{i\omega t}\langle S^\alpha_{\bf k}(t) S^\beta_{-\bf k}(0)\rangle\ , 
\end{eqnarray} 
which is directly related to the spin Green's functions. 
The standard derivation of the single-magnon contribution to the  
transverse component of the dynamical structure factor  
${\cal S}^{+-}({\bf k},\omega)$  
at $T=0$ gives 
\begin{eqnarray} 
\label{S_perp} 
&&{\cal S}^{+-}({\bf k},\omega)= 
\pi\, S\, (u_{\bf k}+v_{\bf k})^2 \nonumber\\ 
&&\phantom{{\cal S}^{+-}({\bf k},\omega)=} 
\times \left[ A^{11}_{\bf k}(\omega)+  \ 
A^{22}_{\bf k}(\omega)+ 2 A^{12}_{\bf k}(\omega)\right]\ , 
\end{eqnarray} 
where the kinematic ($\omega$-independent) form-factor 
$(u_{\bf k}+v_{\bf k})^2=(1-\gamma_{\bf k})/\omega_{\bf k}$ is 
proportional 
to $k$ close to the ``nuclear'' reciprocal lattice point ${\bf K}=0$ 
and is $\sim 1/k$ close to the ``magnetic'' ${\bf Q}=(\pi,\pi)$ point.
It thus enhances the signal close to the AF ordering vector  
and suppresses it close to the zone center. 
Note that the diagonal parts of the Green's function are symmetric 
and off-diagonal parts are asymmetric  
with respect to the transformation ${\bf k} \rightarrow {\bf k}+{\bf Q}$ 
(since $G^{12}\sim\Sigma^{12}$ and  
$\Sigma^{12}_{{\bf k}+{\bf Q}}=-\Sigma^{12}_{\bf k}$). 
Therefore, the sum of the spectral functions in the bracket in 
Eq. (\ref{S_perp}) is, generally speaking,  
different in the magnetic and nuclear parts of the Brillouin zone. 
At $T>0$ above expression (\ref{S_perp}) is modified by the factor
$[1+n_B(\omega)]$, where $n_B(\omega)=[e^{\omega/T}-1]^{-1}$ is the
Bose distribution function.  
  
The density of states associated with magnetic excitations is 
straightforwardly related to the magnon Green's function (\ref{G_full}) 
and is given by  
\begin{eqnarray} 
\label{N_w} 
N(\omega)=\sum_{\bf k}\left[A^{11}_{\bf k}(\omega)+ A^{22}_{\bf 
k}(\omega)\right]\ . 
\end{eqnarray} 
The magnetic specific heat is then given by 
\begin{eqnarray} 
\label{C_T} 
C_M(T)=\frac{1}{T^2}\int_0^1 d\omega\, N(\omega)\, \omega^2\, 
\left[ 
n_B(\omega)^2+n_B(\omega)\right]\ , 
\end{eqnarray} 
where $\omega$ and  $T$ are in the units of $\Omega_0=4SJ$. 
 
The static properties of the system, such as staggered magnetization in 
the ordered phase, N\`eel temperature, and 2D correlation length in 
the paramagnetic phase, are calculated from the spin-wave expression 
of the averaged on-site magnetic moment: 
\begin{eqnarray} 
\label{Sz} 
&&|\langle S^z_i\rangle|=S-\frac{1}{2}\sum_{\bf 
k}\left(\frac{1}{\omega _{\bf k}}-1\right)\\ 
&&\phantom{|\langle S^z_i\rangle|=} 
-\sum_{\bf k}\frac{1}{\omega _{\bf k}}\left[\langle 
\alpha _{\bf k}^\dag\alpha _{\bf k}\rangle -\gamma _{\bf k} 
\langle \alpha _{\bf k}^\dag\beta_{\bf k}^\dag\rangle\right] 
\nonumber \, 
\end{eqnarray} 
where bosonic averages can be expressed through the spectral 
functions (\ref{A_ij}) as: 
\begin{eqnarray} 
\label{averages} 
&&\langle\alpha _{\bf k}^\dag\alpha _{\bf k}\rangle=\int_{-\infty 
}^\infty d\omega \, n_B(\omega)\, A^{11}_{R,{\bf k}}(\omega )\ ,\\ 
&&\langle\alpha _{\bf k}^\dag\beta^\dag_{\bf k}\rangle=\int_{-\infty 
}^\infty d\omega \, n_B(\omega)\, A^{12}_{R,{\bf k}}(\omega )\ ,\nonumber 
\end{eqnarray} 
which implicitly depend on impurity concentration $x$, index $R$ 
stands for retarded.  
 
In the ordered phase these expressions (\ref{Sz}), (\ref{averages}) 
provide us with the concentration and temperature dependence of the 
averaged staggered magnetization $M(x,T)$.  
The same expressions with the 
condition $\langle S^z\rangle(x,T)=0$ define the mean-field equation 
on the N\'eel temperature as a function of $x$. 
In both cases, when $T\neq 0$ the 3D form of the spin-wave dispersion 
is to be used in Eq. (\ref{Sz}). 
In the paramagnetic phase ($T>T_N$) Eq. (\ref{Sz}) should be modified by 
$\gamma_{\bf k}\rightarrow \eta\gamma_{\bf k}$ and $\omega_{\bf 
k}\rightarrow \sqrt{1-\eta^2\gamma_{\bf k}^2}$. Then, in the framework 
of the modified spin-wave theory \cite{Takahashi}, equation $\langle 
S_i^z\rangle(x,T,\eta)=0$ is a constraint which represents   
a self-consistent equation on the gap $\sqrt{1-\eta^2}$. This, in turn,  
defines the 2D correlation length $\xi_{2D}$ as a function of $x$ and $T$. 

\section{Dynamic and thermodynamic properties} 
\label{dynamic} 

In this Section we consider in detail the structure of the spectral
functions of the Green's function Eq. (\ref{G_full}),
Fig. \ref{fig_1}(b), with self-energies given by Eqs. 
(\ref{Sigma_1})-(\ref{sigma_d}). We calculate the dynamical structure
factor ${\cal S}({\bf k},\omega)$, spin-wave density of states
$N(\omega)$, and low-$T$ magnetic specific heat $C_M(T)$.
We consider the long-wavelength,
low-energy limit of the problem and obtain analytical results for the
low-energy ${\cal S}({\bf k},\omega)$ and 
$N(\omega)$ and the low-temperature $C_M(T)$.
We recall here that all wave-vectors are in units of inverse lattice
spacing $1/a$ and all energies are in units of $\Omega_0=4SJ$.

We consider the low-energy form of the Green's functions
first. At low energies $\omega, \omega_{\bf k}\ll 1$ self-energies
(\ref{sigma_s})-(\ref{sigma_d}) are given by:
\begin{eqnarray}
\label{sigma_le}
&&\Sigma_{\bf k}^{11}(\omega)= x\omega_{\bf k}\bigl[
\rho(\omega)+2-\pi/2\bigr]-x\omega+
{\cal O}(\omega_{\bf k}\omega^2\rho^3)\ ,
\nonumber\\
&&\Sigma_{\bf k}^{12}(\omega)= x\omega_{\bf k}\gamma_{\bf k}\bigl[
\rho(\omega)+\pi/2\bigr]+{\cal O}(\omega_{\bf k}\omega^2\rho^3)\ , \\
&&\mbox{with}\ \ \ \rho(\omega)\simeq (2/\pi)\ln|\omega/4|-i\ ,
\nonumber
\end{eqnarray}
which includes contributions from $s$- and $p$-wave scattering,
$d$-wave part is of higher order 
$\Sigma_d\simeq {\cal O}(\omega_{\bf k}^3)$, $\omega_{\bf k}\simeq
k/\sqrt{2}$. The importance of the projection of the unphysical states 
can be demonstrated one more time by comparison of the above
expressions with the ``unprojected'' ($H_z=0$) form of the self-energy:
\begin{eqnarray}
\label{sigma_le_H0}
\Sigma_{\bf k}^{11}(\omega)= x\omega_{\bf k}\bigl[
\rho(\omega)-\pi/2\bigr]+x\omega_{\bf k}^2/\omega+
{\cal O}(\omega_{\bf k}\omega^2\rho^3)\ ,
\end{eqnarray}
which possesses an $\omega=0$ singularity. Noteworthy, the ``physical''
part of the expression containing the logarithm is not related to the
unphysical states and stays intact under the projection. As it was
noted before ``on-shell'', $\omega=\omega_{\bf k}$, the self-energies 
(\ref{sigma_le}) and (\ref{sigma_le_H0}) coincide \cite{Wan}.
The off-diagonal $\Sigma_{\bf k}^{12}(\omega)$ is the same in both
cases. It is also useful to note that the first-order Born
approximation to the scattering problem would give a very different 
result
\begin{eqnarray}
\label{sigma_le_Born}
\Sigma_{\bf k}^{11,Born}(\omega)=- 2x\omega_{\bf k}\ , \ \ \ 
\Sigma_{\bf k}^{12,Born}(\omega)=0 \ ,
\end{eqnarray}
with imaginary part of the self-energy being $\sim{\cal O}(xk\omega^2)$.

One can see that along with the ``normal'' softening  
and weak damping the full $T$-matrix consideration gives 
non-linear dispersion term and damping $|\tilde{\gamma}_{\bf
k}|/\omega_{\bf k}\simeq x$ which is only parametrically
small with respect to the bare spectrum. 
A perturbative ``on-shell'' pole equation gives
\begin{eqnarray}
\label{pert_pole}
&&\tilde{\omega}_{\bf k}+i\tilde{\gamma}_{\bf
k}=\omega_{\bf k}+\Sigma_{\bf k}^{11}(\omega_{\bf k})\ ,\nonumber \\
&&\tilde{\omega}_{\bf k}=\omega_{\bf k}\bigg(1-x(\pi/2-1)+
\frac{2x}{\pi}\ln|\omega_{\bf k}/4|\bigg) \ , \\
&&\tilde{\gamma}_{\bf k}=-x\omega_{\bf k}\ , \nonumber
\end{eqnarray}
which already shows that the spin-wave velocity in the effective medium
is not well defined since the bracket in Eq. (\ref{pert_pole}) depends on
${\bf k}$. Moreover, the renormalization of the real part of the spectrum
is dominated by the $\ln|\omega|$ term at low frequencies and the
bracket vanishes at some wave-vector
\begin{eqnarray}
\label{lambda}
k_c^{-1}\sim\exp(\pi/2x)\ .
\end{eqnarray}
Because of that one can naively suggest a vanishing of the spectrum 
\cite{Wan} and an instability of the ground state towards some new
phase. Such an instability is, of course, just a signature of the
breakdown of the perturbation theory. One has to sum up all the 
``dangerous'' terms
using Belyaev-Dyson equation Fig. \ref{fig_1}(b)
and Eq. (\ref{G_full}) and analyze the spectral functions
(\ref{A_ij}).

The low-energy, long-wavelength form of the Green's functions
Eq. (\ref{G_full}) with self-energies from Eq. (\ref{sigma_le}) is
\begin{eqnarray} 
\label{G_le} 
&&G^{11}_{\bf k}(\omega)=G^{22}_{\bf k}(-\omega)
\simeq\frac{\tilde\omega+\omega_{\bf k}
\bigg(1+x[\rho(\omega)+2-\pi/2]\bigg)}
{\tilde\omega^2 -\omega_{\bf k}^2\bigg(1+x[2\rho(\omega)+4-\pi]\bigg)}
\ , \nonumber\\
&&G^{12}_{\bf k}(\omega)\simeq-
\frac{x\gamma_{\bf k}\omega_{\bf k}[\rho(\omega)+\pi/2]}
{\tilde\omega^2 -\omega_{\bf
k}^2\bigg(1+x[2\rho(\omega)+4-\pi]\bigg)}\ ,
\end{eqnarray} 
where $\tilde\omega=\omega(1+x)$ and $\rho(\omega)$ is defined in
(\ref{sigma_le}). 

The diagonal spectral function in the same limit can be then written as:
\begin{eqnarray} 
\label{A_le} 
A^{11}_{\bf k}(\omega)\simeq\frac{1}{\pi}
\frac{x\omega_{\bf k}\big(\tilde\omega+\omega_{\bf k}\big)^2}
{\big(\tilde\omega^2 -\omega_{\bf k}^2 a(\omega)\big)^2+
\big(2x\omega_{\bf k}^2\big)^2}\ , 
\end{eqnarray} 
where we make use of imaginary part of the self-energies being
Im$\Sigma^{ij}_{\bf k}(\omega)\simeq -x\omega_{\bf k}$, and introduce a
``stretching factor'' 
\begin{eqnarray} 
\label{aw} 
a(\omega)=1+x\big(\frac{4}{\pi}\ln|\omega/4|
+4-\pi\big)\ . 
\end{eqnarray} 
The energy at which this factor vanishes defines the
 disorder-induced energy scale 
\begin{eqnarray} 
\label{w_0} 
\omega_0\sim \exp\left(-\frac{\pi}{4x}\right)
\end{eqnarray} 
below which the spectrum is overdamped. 

A more detailed analysis of Eq. (\ref{A_le}) gives the following picture.
At the wave-vectors much larger than $\omega_0$ ($\omega_{\bf
k}\gg\omega_0$), that is at the wavelengths shorter than a characteristic
length $\ell\sim e^{-\pi/4x}$, the spectral function has three
distinct regions in $\omega$.
First, is a vicinity of a quasiparticle peak,
$\omega\approx\tilde\omega_{\bf k}$:
\begin{eqnarray} 
\label{A_le_1} 
A^{11}_{\bf k}(\omega)\approx\frac{2\omega_{\bf k}}{\pi}
\frac{2x\omega_{\bf k}^2}
{\big(\omega^2 -\tilde\omega_{\bf k}^2 \big)^2+
\big(2x\omega_{\bf k}^2\big)^2}\ , 
\end{eqnarray} 
where the spectrum has a regular Lorentzian form with 
the pole at $\tilde{\omega}_{\bf k}$ and width $\tilde\gamma_{\bf k}$
given by the perturbative result Eq. (\ref{pert_pole}).
Second, the intermediate range of energies, 
$\omega_0<\omega\ll\tilde\omega_{\bf k}$, where the ``stretching
factor'' is not too close to zero:
\begin{eqnarray} 
\label{A_le_2} 
A^{11}_{\bf k}(\omega)\approx\frac{1}{\pi}
\frac{x}{\omega_{\bf k}}\frac{1}
{a(\omega)^2+4x^2}\approx\frac{1}{\pi}
\frac{x}{\omega_{\bf k}}\cdot const\ ,
\end{eqnarray} 
one can approximate $a(\omega)$ by a constant since its dependence on
$\omega$ is weak in this range. One can see that the spectral function
in this 
region is independent of $\omega$ and corresponds to an almost flat,
shallow ($\sim x$) background of states. 
Third, the vicinity of a ``localization peak'', 
$\omega\approx\omega_0$:
\begin{eqnarray} 
\label{A_le_3} 
&&A^{11}_{\bf k}(\omega)\approx\frac{1}{4\pi}
\frac{1}{x\omega_{\bf k}}\ \ \ \ \mbox{at}\ \ \ \omega=\omega_0\ ,\\
&&A^{11}_{\bf k}(\omega)\approx\frac{\pi}{16}
\frac{1}{\omega_{\bf k}}\frac{1}{x\ln^2|\omega|}\ \ \ \ \mbox{at}\ \ \ 
\omega\ll\omega_0\ ,
\nonumber 
\end{eqnarray} 
where the spectral function rises sharply from the shallow background 
states $\sim x$ (\ref{A_le_2}) 
to a peak of the height $\sim 1/x$ and then
vanishes in a singular fashion as $\omega$ approaches zero. 
Note that this peak is non-Lorentian and its position ($\omega=\omega_0$) 
is independent of the value of ${\bf k}$.

Thus, besides the lack of the Lorentz invariance of the quasiparticle
part of the spectrum Eq. (\ref{pert_pole}), every ${\bf k}$-mode
redistributes some of its weight from the energy $\sim \omega_{\bf k}$ 
to a flat background of states between 
$\tilde\omega_{\bf k}$ and $\omega_0$ and
to a peak at $\omega=\omega_0$. 
Such a behavior is similar to the other problems of linearly
dispersive excitations in the presence of disorder in two dimensions
and should be interpreted as the signature of localization
\cite{lee,Sheng}. Then the characteristic length 
\begin{eqnarray} 
\label{l_0} 
\ell\sim \exp\left(\frac{\pi}{4x}\right)
\end{eqnarray} 
is to be understood as a localization length of the spin-waves in our
problem.  

The truly intriguing question is what is happening at the wavelengths
of the order of $\ell$ and beyond. In our approach
for $k \alt \ell^{-1}$ the quasiparticle and localization peaks
merge into a broad incoherent peak that disperses in the momentum
space. One can see that at $k \sim \omega_0 < \ell^{-1}$  factor
$a(\omega)$ is negative and the ``pole'' in Eq. (\ref{A_le}) becomes
pure imaginary. However, since $a(\omega)$ is $\omega$ dependent
this peak is non-Lorentzian and thus can not be associated with the 
``simple'' diffusive mode. Thus, we observe an 
overdamped,  non-Lorentzian
diffuse-like excitation with a characteristic width of the order
of $\omega_{\bf k}$ and peak position roughly at $\omega \alt\omega_{\bf k}$. 
We have to remark here that the nature of the states at the wavelength
above the localization length might be beyond the ability of our
approach and the proper description of them may require a different,
non-perturbative type of study. 

Thus, the structure of the spectral function we discuss above 
demonstrates an unusual, non-hydrodynamic type of behavior of the
spin-excitation spectrum of a  diluted 2D AF. The strong influence of
disorder in the low-energy excitations in 2D results in the failure 
of the averaging procedure, which effectively restores translational
invariance, to recover the long-wavelength excitation
spectrum of this effective medium. Already at 
the energies much larger than the disorder-induced scale
$\omega\sim k\gg\omega_0$
one finds a departure from the hydrodynamics:
while the ``quasiparticle'' excitation can be found, it does not
disperse linearly with ${\bf k}$ and its damping is neither hydrodynamic
nor quasiparticle-like. More importantly,
{\it above} the characteristic wavelength $\ell$ no hydrodynamic
description of excitations is possible. The low-frequency modes do
exist in some form but they cannot be classified in terms of an
effective wave-vector and thus the long-wavelength propagation
is entirely diffusive. 

In addition, the spectra at $k\gg\omega_0$ are not exhausted by the
quasiparticle peak. They also consist of the background of localized
states and a localization peak described in Eqs. (\ref{A_le_2}),
(\ref{A_le_3}). 

\begin{figure}
\unitlength 1cm
\epsfxsize=7 cm
\begin{picture}(7,6.6)
\put(-0.2,0){\rotate[r]{\epsffile{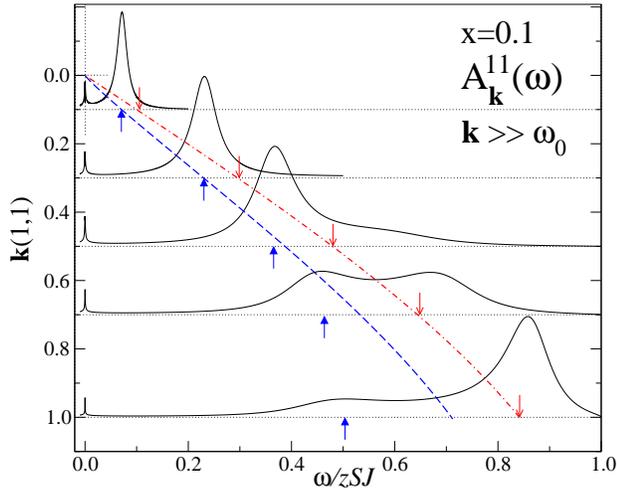}}}
\end{picture}
\caption{The spectral function $A^{11}_{\bf k}(\omega)$ 
for the wave-vectors ${\bf k}=0.1, 0.3, 0.5, 0.7,$ and $1.0$, 
all $\gg\omega_0$  along the $(1,1)$ direction, ${\bf k}$ is in units 
$1/a$. Dashed-dotted line is the bare spin-wave energy,
arrows pointing down are the positions of original delta-function
peaks. Dashed line is the renormalized spin-wave dispersion
Eq. (\ref{pert_pole}), arrows pointing up show the actual
positions of the peaks for selected wave-vectors.
 $A^{11}_{\bf k}(\omega)$ 
for each ${\bf k}$ is normalized to fit the picture.}
\label{fig_2}
\end{figure}

\begin{figure}
\unitlength 1cm
\epsfxsize=7 cm
\begin{picture}(7,6.6)
\put(-0.2,0){\rotate[r]{\epsffile{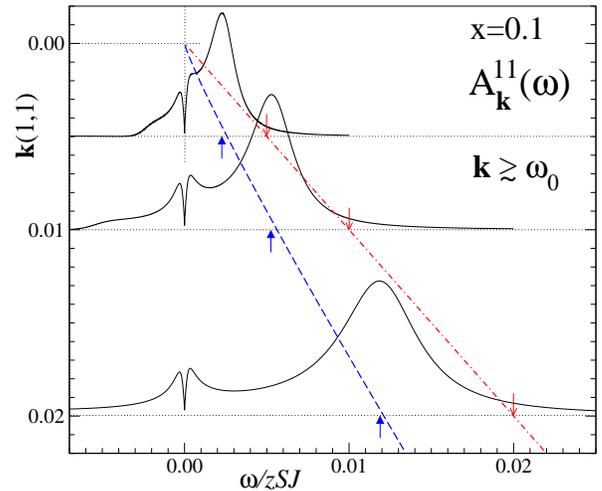}}}
\end{picture}
\caption{The spectral function $A^{11}_{\bf k}(\omega)$ 
for the wave-vectors ${\bf k}=0.005, 0.01,$ and $0.02$
along the $(1,1)$ direction, ${\bf k}=0.005$ is of the order of 
$\omega_0$. Dashed-dotted line, dashed line, and arrows are as in
Fig. \ref{fig_2}.  $A^{11}_{\bf k}(\omega)$ 
for each ${\bf k}$ is normalized to fit the picture.} 
\label{fig_3}
\end{figure}

\begin{figure}
\unitlength 1cm
\epsfxsize=7 cm
\begin{picture}(7,6.6)
\put(-0.2,0){\rotate[r]{\epsffile{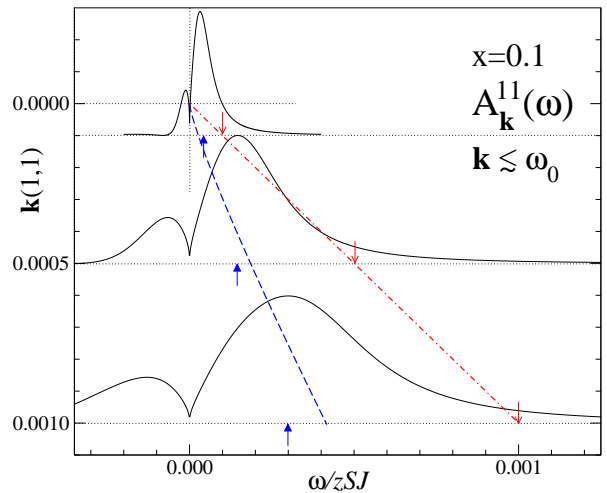}}}
\end{picture}
\caption{The spectral function $A^{11}_{\bf k}(\omega)$ 
for the wave-vectors ${\bf k}=0.0001, 0.0005,$ and $0.001$,
all $<\omega_0$, along the $(1,1)$ direction.
Dashed-dotted line, dashed line, and arrows are as in
Fig. \ref{fig_2}.  $A^{11}_{\bf k}(\omega)$ 
for each ${\bf k}$ is normalized
to fit the picture.} 
\label{fig_4}
\end{figure}

The spectral function $A^{11}_{\bf k}(\omega)$ obtained from the
``full'' expressions for the Green's function (\ref{G_full}) and
self-energies (\ref{Sigma_1})-(\ref{sigma_d})
without taking the low-energy limit is shown in
Figs. \ref{fig_2}-\ref{fig_4} for a number of wave-vectors along the
$(1,1)$ direction of the Brillouin zone for a representative value of
impurity concentration $x=0.1$. The purpose of these pictures is to
demonstrate the features we have discussed using the long-wavelength
form of $A^{11}_{\bf k}(\omega)$. 
The amplitude of each $A_{\bf k}(\omega)$ curve is normalized
to fit the picture and therefore the relative heights of the curves
bear no meaning. These figures also show the bare spin-wave energy
(dashed-dotted line) with arrows pointing down showing the positions
of ``unperturbed'' delta-function peaks. Dashed line corresponds to
the perturbative renormalized spin-wave dispersion,
Eq. (\ref{pert_pole}), while arrows pointing up show the actual
positions of the peaks for selected wave-vectors.
The figures show the spectral function within the different ranges
of ${\bf k}$ relative to $\omega_0$, ${\bf k}\gg\omega_0$, ${\bf
k}\agt\omega_0$,  and ${\bf k}\alt\omega_0$, respectively. 
The latter can be calculated
using Eq. (\ref{w_0}) which gives $\omega_0(x=0.1)\sim 10^{-3}$.
\noindent
\noindent

Fig. \ref{fig_2} shows the spectral function $A^{11}_{\bf k}(\omega)$
for the wave-vectors ${\bf k}=0.1, 0.3, 0.5, 0.7,$ and $1.0$ along the 
$(1,1)$ direction (${\bf k}$ is in units of $1/a$, so that the corner of
the Brillouin zone is $(\pi,\pi)$). One can see that the quasiparticle
peak follows the renormalized spin-wave dispersion (\ref{pert_pole}) 
at low ${\bf k}$
very closely. At higher values of ${\bf k}$ the higher energy sub-band
develops and the spectrum evolves into a ``camel''-like structure
discussed extensively in Ref. \cite{kampf}. 
The origin of this high-energy 
structure is in the presence of the high-energy resonance state 
($\omega_{res}\simeq J$) around impurity \cite{CB} which is 
unrelated to the low-energy physics of the system. 
Since our low-energy consideration does not take this high-energy
feature into account the position of the lower peak in
this structure deviates from the long-wavelength dispersion
(\ref{pert_pole}) at larger ${\bf k}$. 
The low-energy localization peak and background are already noticeable  
in Fig. \ref{fig_2} despite the high-energy scale.

Fig. \ref{fig_3} shows the spectral function $A^{11}_{\bf k}(\omega)$
for the wave-vectors ${\bf k}=0.005, 0.01,$ and $0.02$ with the
smallest  wave-vector being of the order of $\omega_0$.
One can clearly see the features we have discussed in
Eqs. (\ref{A_le_1})-(\ref{A_le_3}): the broadened quasiparticle peak,
the localization peak, and the states between them.
The quasiparticle peak continue to follow the renormalized spin-wave 
dispersion (\ref{pert_pole}).
As the ${\bf k}$ decreases all the mentioned structures merge.
\noindent
\noindent
\noindent
\noindent

Fig. \ref{fig_4} shows the spectral function $A^{11}_{\bf k}(\omega)$
for the wave-vectors ${\bf k}=0.0001, 0.0005,$ and $0.001$,
all are smaller than $\omega_0$.
As we discussed above, the quasiparticle and localization peaks
merge and give a broad, overdamped, non-Lorentzian diffusive peak.
In other words, one may not represent the Green's function in this
region as a sum of coherent and incoherent contributions
$G^{coh}_{\bf k}(\omega)+G^{incoh}_{\bf k}(\omega)$, it seems that
only the second part survives.  
The peak position deviates from the perturbative 
renormalized spin-wave dispersion (\ref{pert_pole}) and thus indicate
the region where the perturbation theory breaks down.

The off-diagonal spectral function $A^{12}_{\bf k}(\omega)$ should
possess features similar to the one of the diagonal spectral function.
The low-energy, long-wavelength form of $A^{12}_{\bf k}(\omega)$ is 
given by
\begin{eqnarray} 
\label{Aoff_le} 
A^{12}_{\bf k}(\omega)\simeq\frac{1}{\pi}
\frac{x\gamma_{\bf k}\omega_{\bf k}\big(\omega_{\bf k}^2 \, b(x)
-\tilde\omega^2\big)}
{\big(\tilde\omega^2 -\omega_{\bf k}^2 a(\omega)\big)^2 +
\big(2x\omega_{\bf k}^2\big)^2}\ , 
\end{eqnarray} 
where $b(x)=\big[1-2x(\pi-2)\big]$. Note that 
$A^{12}_{\bf k}(\omega)$ is not a positively defined function, 
it changes sign as a function of $\omega$ at $\omega=\omega_{\bf k}
\sqrt{b(x)}/(1+x)$. Another important difference from $A^{11}_{\bf
k}(\omega)$ is that $A^{12}_{\bf k}(\omega)$ is odd under
the transformation ${\bf k} \rightarrow {\bf k}+{\bf Q}$ and thus has
opposite sign in the first and second magnetic Brillouin zones.

A detailed analysis of $A^{12}_{\bf k}(\omega)$
in different regions of $\omega_{\bf k}$ and $\omega$ 
shows that in the vicinity of a quasiparticle peak $A^{12}_{\bf
k}(\omega)$ has an additional smallness of order $x$ 
in comparison with $A^{11}_{\bf k}(\omega)$, but it is of the same
order 
in the ``intermediate'' ($\omega<\omega_{\bf k}$) and low-energy
regions where it can be approximated as
\begin{eqnarray} 
\label{Aoff_le_2} 
A^{12}_{\bf k}(\omega)\approx\frac{1}{\pi}
\frac{x\gamma_{\bf k}}{\omega_{\bf k}}\frac{1}
{a(\omega)^2+4x^2}\ ,
\end{eqnarray} 
with the behavior above, at, and below the localization peak 
identical to the one of $A^{11}_{\bf k}(\omega)$, Eqs. (\ref{A_le_2}), 
(\ref{A_le_3}).
\begin{figure}
\unitlength 1cm
\epsfxsize=7 cm
\begin{picture}(7,6.6)
\put(-0.2,0){\rotate[r]{\epsffile{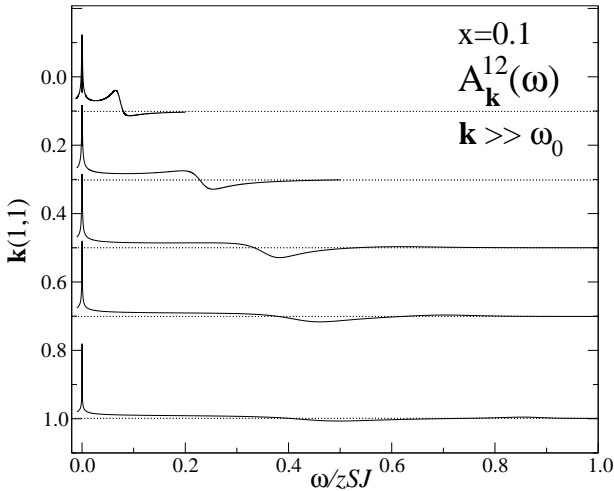}}}
\end{picture}
\caption{The spectral function $A^{12}_{\bf k}(\omega)$ 
for the wave-vectors ${\bf k}=0.1, 0.3, 0.5, 0.7,$ and $1.0$, 
all $\gg\omega_0$  along the $(1,1)$ direction. 
 $A^{12}_{\bf k}(\omega)$ 
for each ${\bf k}$ is normalized to fit the picture.}
\label{fig_5}
\end{figure}
\noindent

Fig. \ref{fig_5} gives an example of the structure of the off-diagonal
spectral function $A^{12}_{\bf k}(\omega)$ obtained from Eqs.
(\ref{G_full}) and (\ref{Sigma_1})-(\ref{sigma_d})
without taking the low-energy limit
for the wave-vectors ${\bf k}=0.1, 0.3, 0.5, 0.7,$ and $1.0$. 
The features discussed in the preceding paragraphs 
such as change of the sign, low-energy localization 
peak, and background states are clearly seen in this spectral
function. 

\begin{figure}
\unitlength 1cm
\epsfxsize=7 cm
\begin{picture}(7,6.6)
\put(-0.2,0){\rotate[r]{\epsffile{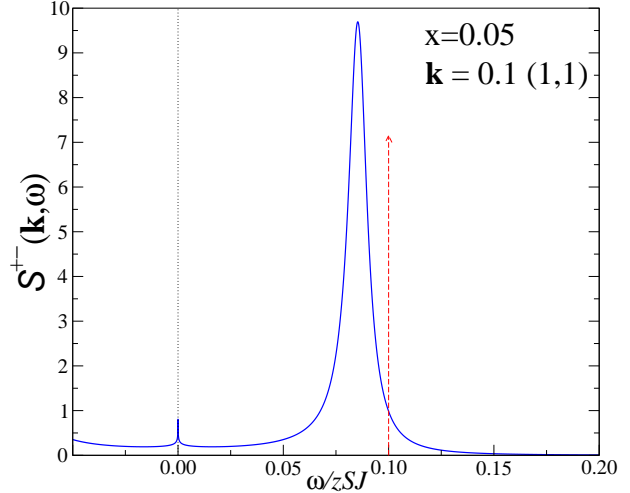}}}
\end{picture}
\caption{Transverse component of the neutron-scattering dynamical 
structure factor  ${\cal S}^{+-}({\bf k},\omega)$, ${\bf k}=0.1$,
$x=0.05$. Long dashed arrow shows the initial position of 
delta-functional peak.}
\label{fig_6}
\end{figure}

The transverse component of the neutron-scattering dynamical 
structure factor  ${\cal S}^{+-}({\bf k},\omega)$
is directly related to the linear combination of the
magnon spectral functions $A^{11}_{\bf k}(\omega)$, 
$A^{22}_{\bf k}(\omega)$(=$A^{11}_{\bf k}(-\omega)$), and 
$A^{12}_{\bf k}(\omega)$ as given by Eq. (\ref{S_perp}).
It therefore must contain all the features of the spectral
functions we discuss here. Fig. \ref{fig_6} shows an example of our
result for ${\cal S}^{+-}({\bf k},\omega)$ v.s. $\omega$ at ${\bf
k}=0.1$, that is in the ``nuclear'' Brillouin zone, for $x=0.05$. Long
dashed arrow shows the initial position of delta-functional peak.
Since $\pm\omega_0$ are very small in this case the localization peak is
seen as a single 
spike at $\omega = 0$, but the flat background of states
is clearly visible below the quasiparticle peak.
\noindent

However, the actual observation of anomalous features of the 
spectra can be complicated because of two reasons. 
First, the structure factor contains a
kinematic form-factor which enhances the spectral function
combination by $\simeq 2/\omega_{\bf k}$ close to ${\bf k}={\bf Q}$ and
suppresses it by $\simeq \omega_{\bf k}/2$ close to ${\bf k}=0$.
Second, as we show below, the sum of the spectral functions entering  
${\cal S}^{+-}({\bf k},\omega)$ is ``less anomalous'' close to 
${\bf k}={\bf Q}$ than at ${\bf k}\rightarrow 0$. Namely, the
quasiparticle part of the spectrum is abnormally broadened and
disperses nonlinearly for both ${\bf k}\rightarrow 0$ and 
${\bf k}\rightarrow {\bf Q}$, while the low-energy localization
features are suppressed in the vicinity of ${\bf Q}$ due to
cancellation between the diagonal and off-diagonal contributions.

One can show explicitly using the low-energy, long-wavelength limit
of the sum of spectral functions $\big[A^{11}_{\bf k}(\omega)+
A^{22}_{\bf k}(\omega)+2 A^{12}_{\bf k}(\omega)\big]$ given by
\begin{eqnarray} 
\label{S_le} 
&&A^{\Sigma}_{\bf k}(\omega)\equiv
\sum_{\alpha\beta=1,2}A^{\alpha\beta}_{\bf k}(\omega)
\simeq\frac{2x\omega_{\bf k}}{\pi}\\
&&\phantom{\sum_{\alpha\beta=1,2}}
\times\frac{\tilde\omega^2(1-\gamma_{\bf
k})+ \omega_{\bf k}^2(1+\gamma_{\bf k})-2x\gamma_{\bf k}\omega_{\bf
k}^2(\pi-2)}
{\big(\tilde\omega^2 -\omega_{\bf k}^2 a(\omega)\big)^2 +
\big(2x\omega_{\bf k}^2\big)^2}\ , \nonumber
\end{eqnarray} 
that aside from the kinematic form-factor the dynamical   
structure factor should be different in the first 
(${\bf k}\rightarrow 0$) 
\begin{eqnarray} 
\label{S_le_1} 
A^{\Sigma}_{\bf k}(\omega)\approx
\frac{1}{\pi}
\frac{4x\omega_{\bf k}^3}
{\big(\tilde\omega^2 -\omega_{\bf k}^2 a(\omega)\big)^2 +
\big(2x\omega_{\bf k}^2\big)^2}\ , \
\end{eqnarray}
and the second (${\bf k}\rightarrow {\bf Q}$) 
magnetic Brillouin zones 
\begin{eqnarray} 
\label{S_le_2} 
A^{\Sigma}_{\bf k}(\omega)\approx
\frac{1}{\pi}
\frac{4x\omega_{\bf k}\tilde\omega^2}
{\big(\tilde\omega^2 -\omega_{\bf k}^2 a(\omega)\big)^2 +
\big(2x\omega_{\bf k}^2\big)^2}\ ,
\end{eqnarray} 
 due to the asymmetry of
$A^{12}_{\bf k}(\omega)$ to ${\bf k}\rightarrow {\bf k}+{\bf Q}$.
One can see that around the quasiparticle peak $\omega\simeq\omega_{\bf
k}$ these expressions are identical and are simply equal to the diagonal
spectral function Eq. (\ref{A_le_1}), but at lower energies for 
${\bf k}\sim{\bf Q}$ the localization features are suppressed by the
factor of $\omega^2$.

This asymmetry is demonstrated in Fig. \ref{fig_7} which shows 
the intensity map of 
${\cal S}^{+-}({\bf k},\omega)\cdot \omega_{\bf k}/(1-\gamma_{\bf
k})= \pi S A^{\Sigma}_{\bf k}(\omega)$, 
that is the structure factor divided by the kinematic form-factor,
in the ${\bf k}-\omega$ plane
across the Brillouin zone in the $(1,1)$ direction from ${\bf k}=0$
to ${\bf k}=(\pi,\pi)$, for $x=0.25$. The higher intensity corresponds
to the higher value of the function. 
One can clearly see all the features of the spectrum described in this
Section: the resonance and its splitting from the dispersive mode at
high energies, the low-energy damped spin-wave mode in both the center
and the corner of
the Brillouin zone, and the asymmetric background of localized states
with the low-energy peak at the bottom. 
The nonlinearity of the quasiparticle mode also seem to be 
quite visible though the actual detection of it or of the
abnormal $k$-dependence 
of the damping can be a challenging experimental problem. 

\noindent
\begin{figure}
\unitlength 1cm
\epsfxsize=9.cm
\begin{picture}(9,6.5)
\end{picture}
\caption{The intensity map of 
$\pi S A^{\Sigma}_{\bf k}(\omega)$, 
in the ${\bf k}-\omega$ plane for ${\bf k}$ from $(0,0)$
to $(\pi,\pi)$ in the $(1,1)$ direction and from $\omega=0$ to
$\omega=1$ for $x=0.25$.}
\label{fig_7}
\end{figure}
\noindent

The density of states of spin-excitations can
be easily calculated using Eqs. (\ref{N_w}) and (\ref{G_full}).
We recall that for the pure 2D system with linear spectrum of 
excitations low-energy density of states is a linear function of
$\omega$ and in our case
\begin{eqnarray} 
\label{N_w1} 
N(\omega)=\frac{2}{\pi}\omega\ .
\end{eqnarray} 
Evidently, the low-energy localized states should strongly affect
$N(\omega)$ and one readily finds the anomalous corrections already on
the level of perturbative analysis of the Green's function. 
If one uses the full $T$-matrix form of the self-energy but expands the
Green's function in $x$: 
\begin{eqnarray} 
\label{G_pert} 
G^{11}_{\bf k}(\omega)\simeq G^{0,11}_{\bf k}(\omega)+
G^{0,11}_{\bf k}(\omega)\Sigma^{11}_{\bf k}(\omega)G^{0,11}_{\bf
k}(\omega) \ ,
\end{eqnarray} 
one immediately gets a constant correction 
\begin{eqnarray} 
\label{N_w2} 
N(\omega)=\frac{2}{\pi}\omega+xC+{\cal
O}(x\omega\ln|\omega|)\ ,
\end{eqnarray} 
which also implies a finite density of states at $\omega =0$.
A more sensible result can be obtained without using $x$-expansion
from the long-wavelength expression for  the spectral function
$A^{11}_{\bf k}(\omega)$ (\ref{A_le}):
\begin{eqnarray} 
\label{N_w3} 
N(\omega)=\frac{2}{\pi}\omega+xC_1/\big[a(\omega)^2+4x^2\big]+{\cal
O}(x\omega\ln|\omega|)\ ,
\end{eqnarray} 
where $a(\omega)$ is the same ``stretching factor'' Eq. (\ref{aw}) 
we used in
Eqs. (\ref{A_le}), (\ref{A_le_2}), (\ref{Aoff_le})-(\ref{S_le_2}). 
At $\omega\gg \omega_0$  $a(\omega)\approx const$ and we are back
to the previous result given by $x$-expansion 
perturbation theory (\ref{N_w2}). 
At $\omega \approx \omega_0$ density of states 
has a peak  of the height $\sim 1/x$ whose origin is evident: 
the low-energy non-dispersive localized states altogether contribute to
it. At  $\omega \ll \omega_0$  the density of states 
vanishes as $N\propto 1/(x\ln|\omega|)^2$.
Such a strong dependence of the result on the degree of approximation
is reminiscent to the dispute over $N(\omega)$ 
for the certain types of disorder in 2D systems with 
linear excitation spectrum \cite{lee,Yashenkin} where different
approaches result in drastically different answers for the low-energy
part of the density of states. 
\noindent
\begin{figure}
\unitlength 1cm
\epsfxsize=7 cm
\begin{picture}(7,6.6)
\put(-0.2,0){\rotate[r]{\epsffile{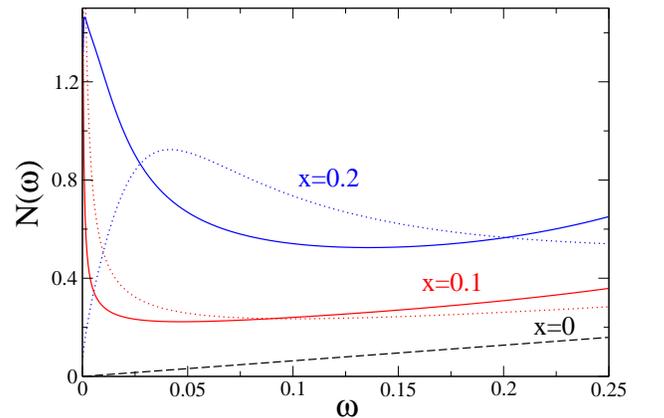}}}
\end{picture}
\caption{Density of states $N(\omega)$ v.s. $\omega$ for 
$x=0$ (pure system, dashed curve), $x=0.1$, and $x=0.2$ (dotted and
solid curves). 
Dotted curves are the long-wavelength result 
Eq. (\ref{N_w3}) with $C_1=4/\pi^{3/2}$.
Solid curves are the
result of numerical integration using the full Green's function
(\ref{G_full}).}
\label{fig_8}
\end{figure}
\noindent

Our Fig. \ref{fig_8} shows the results for the density of states for 
$x=0$ (pure system, dashed curve), $x=0.1$, and $x=0.2$. 
The dotted curves show $N(\omega)$ given by Eq. (\ref{N_w3}) with
$C_1=4/\pi^{3/2}$ which is obtained from a long-wavelength expression
for the spectral function (\ref{A_le}). The solid curves are the
result of numerical integration using the ``full'' Green's function
(\ref{G_full}). While the overall agreement of these curves is very
good there is a significant discrepancy at low energies which has the
following origin. In the long-wavelength limit we regarded the
localization peak at $\omega=\omega_0$ as non-dispersive, whereas at
larger ${\bf k}$, close to the magnetic Brillouin zone boundary, it
does disperse down to $\omega\sim\omega_0^2\sim e^{-\pi/2x}\ll
\omega_0$. This can
be noticed in our Fig. \ref{fig_7} as well. 
As a result of such a dispersion the peak in the density of states at
$\omega_0$ is spread to lower energies. Technically, there is a
term in denominator of the Green's function $\sim x^2\rho(\omega)^2
\omega_{\bf k}^4$, negligible at low ${\bf k}$, 
which leads to such a behavior. 
 Since this term is
of the order of $x^2$ and our approach does not take into
account all such terms we have no certainty on whether it is a
spurious feature or not. As we show below this discrepancy
does not affects any of our conclusions.

In this context it is interesting to note that the constant term in 
the density of states, which is a prominent feature of all three 
``full'', long-wavelength, and  perturbative results,
is directly related to the flat background of states 
below the quasiparticle peak in the spectral function. 
The localization-peak feature of the spectral function is responsible 
for the peak in $N(\omega)$ at low $\omega$.
\noindent
\begin{figure}
\unitlength 1cm
\epsfxsize=7 cm
\begin{picture}(7,6.5)
\put(-0.2,0){\rotate[r]{\epsffile{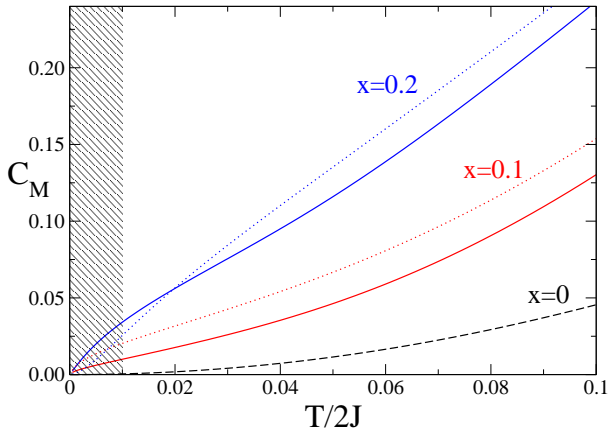}}}
\end{picture}
\caption{$C_M(T)$ v.s. $T$ for the spin-$1/2$ system 
for $x=0$ (dashed curve), $x=0.1$, and $x=0.2$ (dotted and solid
curves). Dotted curves are the long-wavelength results,
solid curves are the result of numerical integration using the 
full Green's function (\ref{G_full}).
The dashed sector shows the 3D crossover temperature region
$T\le\sqrt{J J_\perp}$ for $J_\perp=10^{-4}J$.}
\label{fig_9}
\end{figure}
\noindent

The calculation of magnetic contribution to the specific heat
using results for $N(\omega)$  and Eq. (\ref{C_T}) is straightforward.
For the pure system $C_M(T)\sim T^2$ because of the
two-dimensionality. The anomalous density of states results into a
quasi-linear correction to it. Using the long-wavelength expression
for the density of states we obtain for such a correction:
\begin{eqnarray} 
\label{dC_T} 
\delta C_M(T)\approx \frac{A(x)}{x}\frac{T}{\ln^2|T/\omega_0| + \pi^2/4}
\ ,
\end{eqnarray} 
where $A(x)$ is a weak function of $x$. At $T\gg\omega_0$ ($T$ is also
in units of $\Omega_0$) this gives 
\begin{eqnarray} 
\label{dC_T1} 
\delta C_M(T)\approx xT\cdot const \ . 
\end{eqnarray} 

Fig. \ref{fig_9} shows our results for the magnetic specific heat of
the spin-$1/2$ system ($\Omega_0=2J$) v.s. $T$ for $x=0$ (dashed
curve), $x=0.1$, and $x=0.2$. Dotted curves are the results from the
long-wavelength expression for $N(\omega)$, solid curves are from 
numerical integration using Eqs. (\ref{G_full}) and (\ref{C_T}). 
One can see that the results are very close and point to the same
behavior. The dashed sector shows the temperature region
$T\alt\sqrt{J J_\perp}$ where the crossover to the 3D behavior
(which provides higher powers of $T$ to $C_M$) should occur. We use
the value $J_\perp=10^{-4}J$ characteristic for the cuprates. 

Realistically, this picture should be overlapped with the
phonon contribution to the specific heat. One would 
expect phonons to remain essentially three-dimensional even in 
the layered materials with the characteristic $T^3$ contribution 
to the specific heat at low temperatures and thus be negligible in
comparison with $T^2$ and $xT$ terms. 
However, in the case of cuprates the phonon Debye energy is of the order 
of $400 K$ \cite{Ginzberg} 
which is significantly lower than $2J\simeq 3000 K$. This
makes the phonon part of $C(T)$ to deviate 
from the $T^3$ behavior already 
at about $20 K$, that is around the 3D crossover
temperature for magnetic subsystem. Because of much lower Debye
energy the specific heat in cuprates is dominated by the phonon part
\cite{Ginzberg}. 
Therefore, in order to observe the anomalous quasi-linear contribution
of the localized states to $C(T)$ one needs to use the 
``reference'' material $x=0$ and subtract $C_{x=0}(T)$ from the
results for the systems with $x>0$
(we assume that impurities do not introduce dramatic changes in
the low-energy phonon spectra). Another route is to find a
quasi-2D system with much lower value of $J$ (of the order or less
than Debye energy for phonons) which would allow a direct observation
of $xT$-anomaly from localized states.

The finite value of the inter-plane coupling together with the small
anisotropy gaps leads to the finite value of the ordering temperature
$T_N$ whose dependence on impurity concentration is considered in the
next Session. The effect of the 3D coupling in the dynamic properties,
briefly mentioned above in the context of the specific heat, is the
following. The energy scale introduced by the inter-plane coupling
$\tau=J_\perp/2J$ is $\omega_{3D}=\sqrt{4\tau}$ as seen from
Eqs. (\ref{hat_gamma}), (\ref{wk}) and therefore is rather small for
the realistic systems of interest (for LCO $\omega_{3D}\simeq 0.01$ in
the units of the magnon bandwidth). We show in Appendix \ref{app_B}
that the 3D corrections to the 2D scattering are given by ${\cal
O}(\tau\ln\tau)$ which is truly negligible ($\sim 10^{-4}$ for LCO).
Therefore, the only appreciable correction to the dynamic properties
from 3D coupling is the low-energy cut-off of the logarithmic terms in
the self-energy at $\omega=\omega_{3D}$. As we describe in Appendix
\ref{app_E} 
\begin{eqnarray} 
\label{rho_3D} 
\rho_{3D}(\omega)\simeq\frac{1}{\pi}\ln\left|\frac{\tau}{16}\right|-
i\frac{\omega}{\pi\sqrt{\tau}} \ \ \ \mbox{at} \ \ \omega\ll\omega_{3D}
\ ,
\end{eqnarray} 
that is,  below the 3D energy scale 
the real part is a constant and imaginary part has an extra
power of $\omega$ in comparison with the pure 2D form of
$\rho(\omega)$. $\rho(\omega)$
remains essentially two-dimensional at $\omega >\omega_{3D}$.
Evidently, this proves that the 3D coupling
has little or no effect on the properties of the
spectral functions, dynamical structure factor, or density of
states at $\omega >\omega_{3D}$. 

However, the 3D coupling does affect some of the localization features
in the following way. Below the 3D energy scale the
``stretching factor'' (\ref{aw}) 
saturates at the value $a(\omega_{3D})$ and the
imaginary part of the self-energy acquires an extra power of $\omega$.
In other words, it should be understood as the competition of
disorder-induced and 3D energy scales. Therefore, there are two
regions of $x$. First, when $x$ is small enough $0<x\alt x^*\sim
1/\ln\tau^{-1}$ so that $a(\omega_{3D})>0$. In this region the 
well-defined spin waves can be found deep in the low-${\bf k}$, 
low-$\omega$ region
($k,\omega\ll\omega_{3D}$), similar to the quasi-1D problem \cite{SK}. 
Concentration $x^*$ is defined from the equality of the
energy scales $e^{-\pi/4x}=\sqrt{\tau}$ which gives $x^*\sim
1/\ln\tau^{-1}$.
The localization peak in the spectral function at
$\omega\sim\omega_0\sim Je^{-\pi/4x}$ in the low-$\omega$, ${\bf
k}\gg\omega$ region will be replaced by
\begin{eqnarray} 
\label{A_3D} 
A^{3D,11}_{\bf k}(\omega)\approx\frac{1}{\pi}
\frac{x\omega}{\omega_{\bf k}}\frac{1}
{a(\omega_{3D})^2}\ \ \ \mbox{at} \ \ \omega\ll\omega_{3D}
\ ,
\end{eqnarray} 
which smoothly vanishes as $\omega$ goes to zero instead of showing a
peak. However, the nonlinearity of the spectrum, abnormal damping
of the quasiparticles, and the flat background of the localized 
states below $\tilde{\omega}_{\bf k}$ are all in the 2D-region of ${\bf
k}-\omega$ space ($\omega>\omega_{3D}$) and will remain intact.

Second region is $x\agt x^*$ where $a(\omega_{3D})<0$.
In this region the pole at low-${\bf k}$ and low-$\omega$ 
becomes pure imaginary as in 2D case  
 and the localization peak for low-$\omega$, ${\bf
k}\gg\omega$ reappear above the 3D scale. 
Above the concentration $x^*$ all the low-energy excitations are
incoherent because the 2D disorder-induced energy scale $\omega_0$
(localization length $\ell$) is 
larger (shorter) than the 3D energy scale $\omega_{3D}$ (length scale 
$1/\sqrt{\tau}$) so the spin waves lose their coherence 
before they can propagate in 3D.
A self-consistent calculation is required to determine accurately the 
value of $x^*$ and the details of the 3D to 2D crossover. Our
estimation gives $x^*\sim 0.1-0.2$ for $\tau\sim 10^{-4}$.

Thus, we find that the 3D coupling for the realistic materials will
modify the 2D density of states, structure factor, and specific heat
only at the energies (temperatures) $\omega < \omega_{3D}\simeq
0.01$ and at impurity concentrations $x < x^*\simeq 0.1-0.2$.
The estimated value of the 3D coupling $\tau_c$ which would make $x^*$
larger than the percolation threshold is $\tau_c\sim 0.01$.

 The consideration given above also
applies to the case of small anisotropies introducing gaps in the spectrum 
with a modified $\tau=\tau_{eff}$ accumulating the total effect 
of the gaps and 3D coupling.
It should be noted that the incoherence comes
from the averaging procedure which converts the dissipation
of momentum into the dissipation of the energy. Therefore, the overdamped
excitations should be understood as diffusive.
It is interesting that it requires 2D and ``strong'' disorder to restrict
the number of Euclidean paths for spin waves and 
to break down the description of the problem in terms of an effective
medium. 

\section{Static properties} 
\label{static} 

The static properties such as 
average staggered magnetization $M(x,T)$,
N\'eel temperature $T_N(x)$, and 2D correlation length $\xi(T,x)$ are
considered in this Section.

The average on-site magnetic moment Eq. (\ref{Sz}) for randomly
diluted AF with the averaging over magnetic sites
$M(x)=\sum_i |S^z_i|/N_m$, see \cite{remark1}, can be expressed
through the integral of the spectral functions (\ref{A_ij}) as:
\begin{eqnarray} 
\label{M} 
&&M(x,T)=S-\Delta-\delta M(x,T)\ ,\\
&&\delta M(x,T)=\sum_{\bf k}
\int_{-\infty}^{\infty}\frac{n_B(\omega)\,\mbox{d}\omega}{\omega_{\bf k}}
\big[ A^{11}_{R,{\bf k}}(\omega)-\gamma_{\bf k}A^{12}_{R,{\bf
k}}(\omega)\big]\ , \nonumber
\end{eqnarray} 
where $\Delta=\sum_{\bf k}v_{\bf k}^2\simeq 0.1966$ is the
zero-point spin deviation, $n_B(\omega)=[e^{\omega/T}-1]^{-1}$ is the
Bose distribution function, subscript $R$ denotes retarded.
Note that one should not expect this formula to be valid at large
doping level, $x$ close to $x_p$, since our approach neglects
decoupled clusters and interactions of impurities. However, at not too
large $x$ these effects should be negligible and one expects
Eq. (\ref{M}) to be adequate.
We would also like to note here that our definition of $M(x,T)$ is
physically equivalent to the ``quantum-mechanical factor'' of the
averaged staggered magnetization, the definition 
used in the recent Monte Carlo study
\cite{anders}. In other words, the ``classical'' (``geometrical'')
effect of dilution on magnetization, which simply accounts for the
decrease of the magnetic substance, is multiplicative to the 
quantum effects and is not taken into account in Eq. (\ref{M}).

First we address the question of the presence of explicit divergences
in the integral Eq. (\ref{M}) which would point to the instability of
the long-range order discussed in Refs. \cite{Harris,Wan}.
At $T=0$ $n_B(\omega)=-\theta(-\omega)$ and the impurity-induced quantum
reduction of the magnetization, which can be interpreted as a result
of the ``condensation of magnons'', is given by
\begin{eqnarray} 
\label{dM} 
\delta M(x)=-\sum_{\bf k}\int_{-1}^{0}\frac{\mbox{d}\omega}{\omega_{\bf k}}
\big[A^{11}_{R,{\bf k}}(\omega)-\gamma_{\bf k}A^{12}_{R,{\bf
k}}(\omega)\big]\ , 
\end{eqnarray} 
where we use that the spectral functions are zero outside of the
magnon band $\omega^2>1$. Since the perturbative result
Eq. (\ref{pert_pole}) suggests the instability at small wave-vectors
the long-wavelength expressions for the spectral functions can be used
for our analysis.
From the form of the spectral functions in Eqs. (\ref{A_le}),
(\ref{Aoff_le}) one can readily see that the integral over
$\omega$ is always finite. The integration over ${\bf k}$ is
two-dimensional but has a factor of $1/\omega_{\bf k}$ in the integrand. 
From our expression of the spectral functions in the intermediate
and localization peak energy ranges Eqs. (\ref{A_le_2}),
(\ref{A_le_3}), (\ref{Aoff_le_2}) one may suggest that there is
another $1/\omega_{\bf k}$ in the integrand which would lead to the
logarithmic divergency. However, these expression are obtained by
neglecting $\omega^2$ in comparison with $\omega_{\bf k}^2$ and thus 
are valid at $\omega_{\bf k}\gg \omega$ only. At lower 
${\bf k}$ the convergence of the integral is restored. To show that
more explicitly one can use the $x$-expanded form of the Green's
function Eq. (\ref{G_pert}) for $A^{11}_{R,{\bf k}}(\omega)$
and an equivalent expression for  $A^{12}_{R,{\bf k}}(\omega)$
\begin{eqnarray} 
\label{Goff_pert} 
G^{12}_{\bf k}(\omega)\simeq 
G^{0,22}_{\bf k}(\omega)\Sigma^{12}_{\bf k}(\omega)G^{0,11}_{\bf
k}(\omega) \ .
\end{eqnarray} 
Since all $\Sigma$'s are linear in $x$
this provides an expression for the linear in $x$ term in the
staggered magnetization:
\begin{eqnarray} 
\label{dM1} 
&&\delta M(x)\simeq xB\nonumber\\
&&\phantom{\delta M(x)}
=-\sum_{\bf k}\int_{-1}^{0}\frac{\mbox{d}\omega}
{\pi\omega_{\bf k}}\bigg[\frac{\mbox{Im}
\Sigma^{11}_{R,{\bf k}}(\omega)}{(\omega-\omega_{\bf k})^2}
+\frac{\gamma_{\bf k}\mbox{Im}
\Sigma^{12}_{R,{\bf k}}(\omega)}{\omega^2-\omega_{\bf k}^2}\bigg]
\nonumber\\
&&\phantom{\delta M(x)=}
+\sum_{\bf k}\frac{\gamma_{\bf k}\mbox{Re}
\Sigma^{12}_{\bf k}(\omega_{\bf k})}{2\omega_{\bf k}^2}\ .
\end{eqnarray} 
In the long-wavelength limit this gives
\begin{eqnarray} 
\label{dM2} 
&&\delta M(x)\simeq xB
\simeq\frac{x}{\pi}\sum_{\bf k}
\int_{0}^{1}\mbox{d}\omega\bigg[\frac{1}
{(\omega+\omega_{\bf k})^2}+\frac{1}
{\omega^2-\omega_{\bf k}^2}\bigg]\nonumber\\
&&\phantom{\delta M(x)\simeq xB\simeq}
+\frac{x}{\pi}\sum_{\bf k}
\frac{1}{\omega_{\bf k}}\bigg[ \ln\bigg|\frac{\omega_{\bf
k}}{4}\bigg|-\pi^2/4\bigg] \ ,
\end{eqnarray} 
where the strongest divergency of the integrand is $\ln k\,dk$ and 
all integrals are convergent. 

Numerical integration of the expression in Eq. (\ref{dM1}) 
without the long-wavelength approximation gives the suppression rate of
the staggered magnetization $M(x)\simeq M(0)- Bx$ with $B=0.209(8)$.
For $S=1/2$ it gives the slope of the normalized staggered
magnetization $M(x)/M(0)\simeq
1-Bx/(S-\Delta) \simeq 1-0.691(5)\cdot x$.
It is interesting to note that the second Born approximation
to the impurity scattering
gives three times smaller rate $B_{Born}=0.0725$ showing the necessity 
of the full $T$-matrix treatment of the problem. The 
estimation of $B$, given in the previous study Ref. \cite{Wan} using
$1/z$ approximation for the expression similar to our Eq. (\ref{dM1}), 
provides even smaller $B_{1/z}\simeq 1/z^2=0.0625$ showing yet another
inadequacy of that work.

\begin{figure}
\unitlength 1cm
\epsfxsize=7 cm
\begin{picture}(7,6.6)
\put(-0.2,0){\rotate[r]{\epsffile{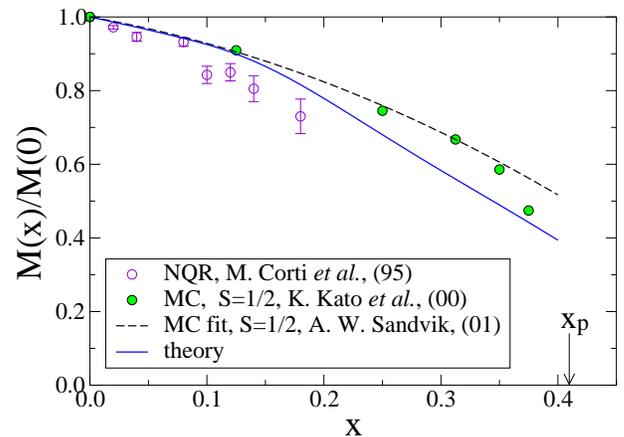}}}
\end{picture}
\caption{Average staggered magnetization v.s. $x$. Our results from
Eq. (\ref{dM}) (solid line), Monte Carlo data (open circles, Ref.
\protect\cite{Kato}), NQR data (filled circles,
Ref. \protect\cite{Corti}), and the fit of Monte Carlo data from
Ref. \protect\cite{anders} are shown.} 
\label{fig_10}
\end{figure}

We have also performed a numerical integration in Eq. (\ref{dM}) for
the impurity-induced reduction of the staggered magnetization 
without $x$-expansion. This yields the results 
presented in Fig. \ref{fig_10} for $S=1/2$ (solid line). 
Monte Carlo data from Ref.
\cite{Kato} (filled circles), 
and NQR data (open circles) from Ref. \cite{Corti} are also shown. 
Note that the original Monte Carlo data of Ref. \cite{Kato}
are normalized by the total number of sites while both NQR and 
our results are averaged over the magnetic sites only. 
In order to extract the same quantity from the Monte Carlo data 
we divided them by the classical probability to find 
a spin-occupied site within the infinite cluster \cite{remark1}. 
A recent Monte Carlo study Ref. \cite{anders} provided an analytical
expression for the fit of the ``quantum-mechanical factor'' in the
magnetization (see the comment after Eq. (\ref{M})) which we plot in
Fig. \ref{fig_10} as well (dashed line). One
can see a very good agreement of our results with numerical data up
to high concentrations. The oxidation of the crystals can be the reason
of a faster decrease of $M(x)$ in NQR data. 

\noindent
\noindent

\begin{figure}
\unitlength 1cm
\epsfxsize=7 cm
\begin{picture}(7,6.6)
\put(-0.2,0){\rotate[r]{\epsffile{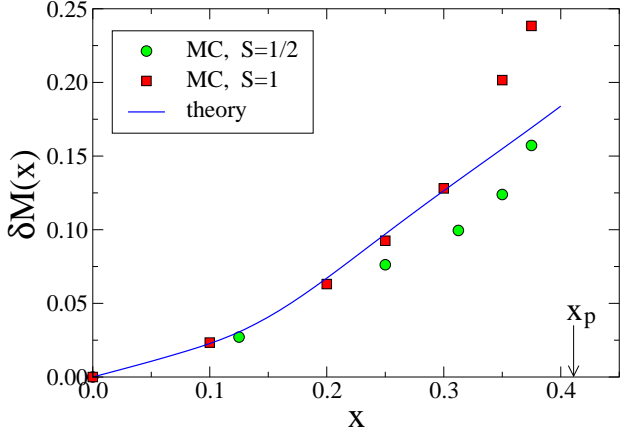}}}
\end{picture}
\caption{The absolute value of $\delta M(x)$ from Eq. (\ref{dM}) (line) and 
Monte Carlo data, Ref.\protect\cite{Kato}, for $S=1/2$ (circles) and $S=1$ 
(squares).}
\label{fig_11}
\end{figure}

The absolute value of impurity-induced quantum fluctuations 
$\delta M(x)$ is independent of $S$ in the linear
spin-wave approximation similar to the quantum reduction of $S$ by
zero-point fluctuations 
$\Delta$. We plot our results for $\delta M(x)$ in Fig. \ref{fig_11}
in order to emphasize the agreement with the MC data for $S=1/2$
(circles) and $S=1$ (squares), which show only weak $S$-dependence. 

It is worth mentioning here that the discrete static
quantities, zero-point spin deviations 
at the neighboring sites around impurities in an AF, were studied 
using spin-wave theory and Green's functions methods since sixties
\cite{Braak} with most recent results obtained 
in Refs. \cite{Mahan}, \cite{Bulut}. Quite
remarkably, these results \cite{Bulut} were found to be in a very good
agreement with the recent 
Monte Carlo studies of 2D $S=1/2$ Heisenberg model with
impurities, Ref. \cite{Sandvik}. Note that while Refs. \cite{Bulut},
\cite{Braak}
were focused on the discrete quantities our results
concern the averaged ones. 

\noindent
\noindent

At $T>0$ Eq. (\ref{M}) for the staggered magnetic moment can be
rewritten separating the quantum, $T=0$, and thermal, $T$-dependent,
parts
\begin{eqnarray} 
\label{M1} 
&&M(x,T)=S-\Delta-\delta M(x)-\delta M^T(x,T)\ ,\\
&&\delta M^T(x,T)=\sum_{\bf k}
\int_{0}^{1}\frac{n_B(\omega)\,\mbox{d}\omega}{\omega_{\bf k}}
\big[ A^{11}_{\bf k}(\omega)+A^{22}_{\bf k}(\omega)\nonumber\\
&&\phantom{\delta M^T(x,T)=\sum_{\bf k}
\int_{0}^{1}\frac{n_B(\omega)\,\mbox{d}\omega}{\omega_{\bf k}}\big[}
-2\gamma_{\bf k}A^{12}_{\bf k}(\omega)\big]\ , \nonumber
\end{eqnarray} 
where $\delta M(x)$ is the zero-temperature part given in
Eq. (\ref{dM}) and we used evident symmetries of the spectral
functions with respect to $\omega\rightarrow -\omega$ and that
$n_B(\omega)=-1-n_B(-\omega)$.

For the true 2D system at $x=0$ and $T>0$ thermal fluctuation destroy
the LRO which manifests itself as a log-divergency of the thermal
correction to the magnetization
\begin{eqnarray} 
\label{dMT} 
&&\delta M^T(0,T)=\sum_{\bf k}
\int_{0}^{1}\frac{n_B(\omega)\,\mbox{d}\omega}{\omega_{\bf k}}
\delta(\omega-\omega_{\bf k})\nonumber\\
&&\phantom{\delta M^T(0,T)}
\simeq \frac{2}{\pi}
\int_{0}^{T}\frac{T\,\mbox{d}\omega}{\omega}\ ,
\end{eqnarray} 
where we use that  $n_B(\omega)\simeq T/\omega$ at $T\ll\omega$.
The 3D coupling provides a cut-off to this divergency in a quasi-2D
problem which yields the finite value of the thermal correction 
\begin{eqnarray} 
\label{dMT1} 
\delta M^T(0,T)\simeq \frac{2}{\pi}
\int_{\sqrt{4\tau}}^{T}n_B(\omega)\,\mbox{d}\omega\simeq \frac{2}{\pi}
T\ln\left|\frac{T}{\sqrt{4\tau}}\right|\ ,
\end{eqnarray} 
and the finite value of the N\'eel temperature whose mean-field value
can be found from the condition $M(0,T)=0=S-\Delta-\delta M^T(0,T)$
which gives
\begin{eqnarray} 
\label{TN1} 
T_N^{MF}\simeq\frac{\pi(S-\Delta)}{\ln\tau^{-1}}\ll 1\ ,
\end{eqnarray} 
in units of $4SJ$. $T_N$ vanishes when $\tau\rightarrow 0$.

One would expect that the thermal part of the staggered
magnetization for the diluted system may possess other divergences,
stronger than the simple log-$\omega$ for the pure system. In fact,
this suggestion is quite natural since the spectrum is not linear and,
therefore, the non-linear corrections must show themselves up. Indeed,
since the correction to the spectrum is $\delta\omega_{\bf k}\sim x
\omega_{\bf k}\ln|\omega|$ (\ref{pert_pole}) one immediately suggests
that the thermal part of the magnetization should acquire a term 
\begin{eqnarray} 
\label{dMT2} 
\sim xT\int \frac{\ln|\omega|\, d\omega}{\omega}\sim xT\ln^2|\omega|\ .
\end{eqnarray} 
However, we show that such anomalous terms from diagonal and
off-diagonal spectral functions cancel each other. As a result, there
is no signature of any new divergency in this quantity caused by the
 anomalies of the spectrum. 

Using the $x$-expanded form for the Green's functions (\ref{G_pert}),
(\ref{Goff_pert}) in the long-wavelength approximation one finds the
diagonal
\begin{eqnarray} 
\label{dMTd} 
&&\delta M_d^T(x,T)=\sum_{\bf k}\frac{1}{\omega_{\bf k}}\langle
\alpha^\dag_{\bf k}\alpha_{\bf k}\rangle^T\simeq\sum_{\bf k}
\int_{0}^{1}\frac{n_B(\omega)\,\mbox{d}\omega}{\pi\omega_{\bf k}}
\\
&&\phantom{\delta M_d^T}
\times\bigg\{\pi\delta(\omega-\omega_{\bf k})\bigg[1-
\frac{\partial\mbox{Re}\Sigma_{\bf k}(\omega_{\bf k})}
{\partial\omega_{\bf k}}\bigg]
-\frac{2x\omega_{\bf k}}{\omega^2-\omega_{\bf
k}^2}\bigg\} \nonumber\\
&&\phantom{\delta M_d^T}
\simeq \frac{2}{\pi}
\int_{0}^{1}n_B(\omega)\,\mbox{d}\omega
\bigg[1-x\bigg(2\rho^\prime(\omega) -\frac{\pi}{2} +1 -\frac{2}{\pi}
\bigg)\bigg] \nonumber \,
\end{eqnarray} 
and off-diagonal
\begin{eqnarray} 
\label{dMTod} 
&&\delta M_{od}^T(x,T)=-\sum_{\bf k}\frac{\gamma_{\bf k}}
{\omega_{\bf k}}\langle \alpha^\dag_{\bf k}\beta^\dag_{\bf k}
\rangle^T\simeq \sum_{\bf k}
\int_{0}^{1}\frac{n_B(\omega)\,\mbox{d}\omega}
{\pi\omega_{\bf k}}\nonumber\\
&&\phantom{\delta M_{od}^T}
\times x \gamma_{\bf k}^2\bigg\{\pi\delta(\omega-\omega_{\bf k})\bigg[
\rho^\prime(\omega_{\bf k})+\frac{\pi}{2}\bigg]+
\frac{2\omega_{\bf k}}{\omega^2-\omega_{\bf
k}^2}\bigg\} \\
&&\phantom{\delta M_{od}^T}
\simeq \frac{2x}{\pi}
\int_{0}^{1}n_B(\omega)\,\mbox{d}\omega
\bigg[2\rho^\prime(\omega) +\frac{\pi}{2} +1\bigg] \nonumber \,
\end{eqnarray} 
parts of the temperature dependent $\delta M^T(x,T)$, where we kept
only ${\cal O}(\ln|\omega|)$ and ${\cal O}(1)$ terms in the integrand,
$\rho^\prime(\omega)\equiv \mbox{Re}\rho(\omega)$, integration by
parts was used in $\delta M_d^T$, superscript $T$ in the averages
means the thermal part.

The total result is 
\begin{eqnarray} 
\label{dMT3} 
\delta M^T(x,T)\simeq \bigg[1+x\bigg(\pi -\frac{2}{\pi}\bigg)\bigg]
\delta M^T(0,T)\ ,
\end{eqnarray} 
which shows that the thermal correction is enhanced by
impurities but there is no new divergency associated with them in
this quantity. Suppression rate of the N\'eel temperature can be
readily obtained from the condition $M(x,T)=0=S-\Delta-\delta
M(x)-\delta M(x,T)$ using Eq. (\ref{dMT3}) which gives:
\begin{eqnarray} 
\label{TN2} 
\frac{T_N(x)}{T_N(0)}\simeq 1-A_s\, x=1-x\bigg(\pi
-\frac{2}{\pi}+\frac{B}{S-\Delta}\bigg) \ .
\end{eqnarray} 
For $S=1/2$ this gives 
$A_{1/2}=3.196(5)$ and for $S=5/2$ it is
$A_{5/2}=2.600(4)$. 
It is important to note that these suppression rates 
point to $x_c(1/2)\simeq 0.31$ and 
$x_c(5/2)\simeq 0.38$,  both below 
$x_p$, so that one may 
suggest that in order to have the phase transition at the classical
percolation threshold the $T_N(x)/T_N(0)$ curves should have a rather
unusual concave form. 

It is interesting to compare our result for the decline rate of
$T_N(x)$  (\ref{TN2}) to
the answers of different approaches to the same problem and to the results
for similar models. A naive mean-field treatment of the impurity
effects as simple renormalization of magnetic coupling
gives $T_N(x)/T_N(0)=1-x$. Application of our
formalism to the Ising limit of the 2D problem gives $T_N(x)/T_N(0)=1-A^I
x$ with $A^I\simeq 1.37$ (see Appendix \ref{app_F}) which is very
close to the RPA answer $A^I_{RPA}= 1.33$ and below the exact answer 
$A^I_{exact}\simeq 1.57$ \cite{McGurn1}. 
For the 2D Ising magnets  $T_N(x)$ v.s. $x$ has a 
more traditional convex form \cite{Cheong}.
Previous result on the suppression rate
of $T_N$ for the 2D Heisenberg model
\cite{McGurn} is $T_N(x)/T_N(0)=1-\pi x$ which is obtained 
using Green's function technique and spin-wave theory in  
approximations very similar to ours. However, Ref. \cite{McGurn} 
misses $-2/\pi$ and neglects $1/S$ terms. 

\noindent
\begin{figure}
\unitlength 1cm
\epsfxsize=7 cm
\begin{picture}(7,6.5)
\put(-0.2,0){\rotate[r]{\epsffile{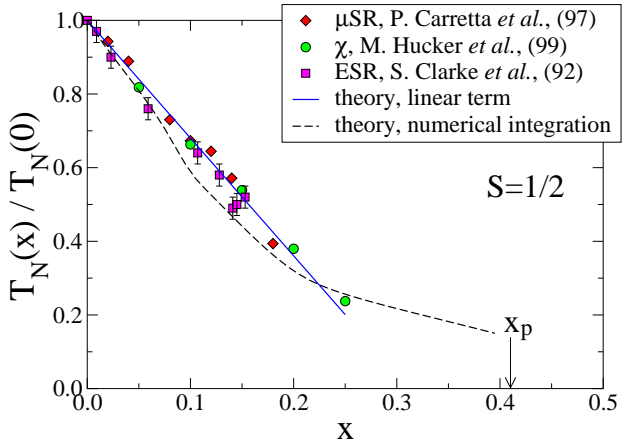}}}
\end{picture}
\caption{$T_N(x)/T_N(0)$ v.s. x for $S=1/2$.
Results of numerical integration in Eq. (\ref{M}) (dashed line),
analytical linear-$x$ slope $(1-A_{1/2}\, x)$ 
Eq. (\ref{TN2}) (solid line),
$\mu$SR (diamonds) \protect\cite{Carretta1} and magnetic
susceptibility  
(circles) \protect\cite{Hucker} data for Zn-doped LCO, and ESR (squares) 
\protect\cite{Clarke} of Zn-doped copper formate
tetrahydrate Cu$_{1-x}$Zn(Mg)$_x$(HCO$_2$)$\cdot$H$_2$O.}
\label{fig_12}
\end{figure}
\noindent

We have also performed 
a numerical integration in Eq. (\ref{M}) and solved an
implicit equation $M(x,T_N)=0$ on $T_N(x)$ numerically. This procedure
requires the finite 3D coupling and the use of quasi-2D form of the
spectral functions. Since the integration involves an additional
dimension and the 3D region is quite narrow the convergence of the
result as a function of number of ${\bf k}, \omega$-points 
at small $x$ can be an issue.
We plot our numerical results for $T_N(x)/T_N(0)$ for the case of
$S=1/2$  in Fig. \ref{fig_12}
together with the analytical slope Eq. (\ref{TN2}) with
$A_{1/2}=3.2$ and experimental
data. Experimental data are obtained by $\mu$SR
\cite{Carretta1} and magnetic susceptibility 
measurements \cite{Hucker} of LCO
systems and by ESR \cite{Clarke} of Zn-doped copper formate
tetrahydrate, a layered quasi-2D AF. One can see that our linear-$x$ 
results agree very well with the experimental data up to a rather high
doping level $x\approx 0.25$.
There is a slight disagreement between our own 
numerical and linear-$x$ analytical results
already at small $x$ which may be connected not only to the
numerical accuracy but to the corrections of the order $\sim
xT_N/J\sim x/\ln\tau^{-1}$.
Note that the linear-$x$ result
is free from such corrections since it is obtained in the
$\tau\rightarrow 0$ limit.

As it is discussed extensively in Ref. \cite{KK} the spin-wave
theory for layered materials is not really adequate at $T\sim T_N$
because of the lack of the kinematic constraints.
When it is applied  to the mean-field equation $M(x,T_N)=0$
it tends to overestimate the absolute 
value of the N\'eel temperature and has some other artifacts such as
$M(T)\sim T_N-T$ at $T\sim T_N$. This may also provide an additional
$x$-dependence in our numerical values of $T_N(x)/T_N(0)$.
\noindent
\begin{figure}
\unitlength 1cm
\epsfxsize=7 cm
\begin{picture}(7,6.5)
\put(-0.2,0){\rotate[r]{\epsffile{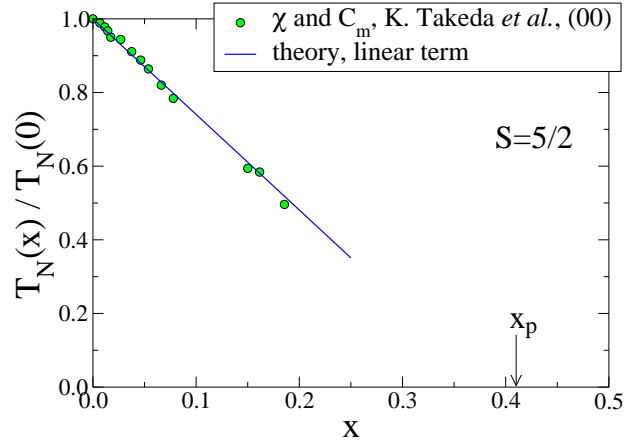}}}
\end{picture}
\caption{$T_N(x)/T_N(0)$ v.s. x for $S=5/2$.
Analytical linear-$x$ results $(1-A_{5/2}\, x)$ 
Eq. (\ref{TN2}) (solid line),
and (circles) \protect\cite{Takeda}
magnetic susceptibility, specific heat data for
Mn$_{1-x}$Zn(Mg,Cd)$_x$(HCO$_2$)$_2$2(NH$_2$)$_2$CO.}
\label{fig_13}
\end{figure}
\noindent

Fig. \ref{fig_13} shows our analytical slope
for $T_N(x)/T_N(0)$ Eq. (\ref{TN2}) 
with $A_{5/2}=2.6$ for the case of $S=5/2$ and 
 experimental data from magnetic susceptibility
and specific heat measurements of 
Mn$_{1-x}$Zn(Mg,Cd)$_x$(HCO$_2$)$_2$2(NH$_2$)$_2$CO, a 
layered $S=5/2$
material, \cite{Takeda}. One can see that while the scattering of
experimental points seem to be smaller than in $S=1/2$ case the 
linear-$x$ result fits them very closely up to $x=0.2$. 
The older $T_N(x)/T_N(0)$ data for the more traditional $S=5/2$ material 
K$_{2}$Mn$_{1-x}$Mg$_{x}$F$_{4}$ \cite{from_Cheong} show a big
scattering of the data which allows almost any reasonable fit
\cite{Cheong}. 

It is worth mentioning that the numerical results 
of our approach for $T_N(x)$
bends inward  at larger values of $x$  
and show the above mentioned concave form, which has been
recently observed experimentally for LCO compounds 
\cite{Greven} and has been anticipated 
in other works \cite{Takeda}. 
While our approach is certainly not adequate at such high impurity
concentrations and tends to overestimate the value of $T_N(x)$ in
comparison with experiments, it nevertheless points to the same
physics. We interpret 
this behavior as due to localization effects which tend to reduce the 
role of quantum and thermal fluctuations in the destruction of the
long-range order.   

In the paramagnetic phase above the N\'eel ordering temperature the 3D
coupling is irrelevant and the spin fluctuations in the layered AF
system are characterized by the in-plane correlation length $\xi_{2D}$
which is exponentially diverging with $1/T$ as $T\rightarrow 0$. The
correlation length is uniquely determined by the $T=0$ properties of
the system such as spin stiffness constant $\rho_s$.

The correlation length can be derived from the modified spin-wave
theory, as was suggested in Ref. \cite{Takahashi}, 
by introducing a chemical potential for magnons which produces
a gap in the spin-wave dispersion and then by resolving a constraint
$\langle S^z_i\rangle=0$ which defines the correlation length
self-consistently. The result of such calculations at $x=0$ is
\cite{Takahashi} 
\begin{eqnarray} 
\label{xi1} 
\xi(T)\simeq\frac{c}{2T}\exp\left(\frac{2\pi\rho_s}{T}\right)\ .
\end{eqnarray} 
One should bear in mind, however, that while this approach gives the
correct exponential behavior  of  $\xi_{2D}(T)$ it provides a
prefactor equivalent to the one-loop renormalization-group
result \cite{chn,Manousakis}. 
This prefactor must be modified according to the higher
order renormalization-group treatment \cite{HN} which gives
\cite{Antonio1} 
\begin{eqnarray} 
\label{xi2} 
\xi(T)\simeq\frac{e c}{2(4\pi\rho_s+T)}
\exp\left(\frac{2\pi\rho_s}{T}\right)\ .
\end{eqnarray} 
This expression 
shows excellent agreement with experiments and Monte Carlo
data \cite{Kastner,Carretta1,xi_exp}. 
This discrepancy between the results of modified
spin-wave theory and result of more exact, non-perturbative approach
is of the same origin as the overestimation of the $T_N$ by the
mean-field solution of $\langle S^z_i\rangle=0$ equation \cite{KK}.

We generalize the approach of Ref. \cite{Takahashi} for the case of an
AF with impurities and obtain for the constraint:
\begin{eqnarray} 
\label{Sz_xi} 
&&S-\frac{1}{2}\sum_{\bf k}\left(\frac{1}{\omega_{\bf
k}(\eta)}-1\right)\\ 
&&=\sum_{\bf k}\frac{1}{\omega _{\bf k}(\eta)}\left[\langle 
\alpha _{\bf k}^\dag\alpha _{\bf k}\rangle -\gamma _{\bf k} 
\langle \alpha _{\bf k}^\dag\beta_{\bf k}^\dag\rangle\right] 
\nonumber \, 
\end{eqnarray} 
where $\omega _{\bf k}(\eta)=\sqrt{1-\eta^2\gamma_{\bf k}^2}$ and
magnon averages are given by the integrals of the spectral functions
$A^{11}_{\bf k}(\omega)$, $A^{12}_{\bf k}(\omega)$ from
Eqs. (\ref{G_full}), (\ref{A_ij}) in which the gapped form of the
spin-wave spectrum is used. 

We have performed a numerical integration in Eq. (\ref{Sz_xi}) 
and calculated the correlation length
$\xi_{2D}(x,T)=\eta^2/\sqrt{8(1-\eta^2)}$ as a function of $T$ for
several values of $x$. We fit the results of such a numerical
procedure in a wide temperature range
almost exactly with the help of original Takahashi formula,
Eq. (\ref{xi1}), with spin-stiffness $\rho_s(x)$ being 
a free parameter. These
fitting values of $\rho_s(x)/\rho_s(0)$ v.s. $x$ follow closely 
our result for $T_N(x)/T_N(0)$ dependence, Fig. \ref{fig_12}. 
Such a result can be anticipated from
the mean-field picture of the ordering in layered systems. The
transition occurs when the inter-plane coupling is strong enough to
stabilize the LRO in comparison with the thermal fluctuations:
$J_\perp M^2(x)\xi^2(x,T_N(x))\approx T_N$. If the correlation length
preserves its exponential form the dominant part of
the left-hand side comes from $e^{2\pi\rho_s(x)/T_N(x)}$ and 
one immediately arrives to
\begin{eqnarray} 
\label{rho_x} 
\frac{\rho_s(x)}{\rho_s(0)}=\frac{T_N(x)}{T_N(0)}+{\cal
O}(x/\ln\tau^{-1}) \ .
\end{eqnarray} 
Therefore, the important conclusion one can make from our analysis is
that\, ({\it i}) the correlation length should follow the $x=0$ type
of behavior Eq. (\ref{xi2}) with $x$-dependent $\rho_s$, 
at least for not too low $T$ and not too high $x$, ({\it ii})
$\rho_s(x)/\rho_s(0)\simeq T_N(x)/T_N(0)$.

\begin{figure}
\unitlength 1cm
\epsfxsize=7 cm
\begin{picture}(7,6.6)
\put(-0.2,0){\rotate[r]{\epsffile{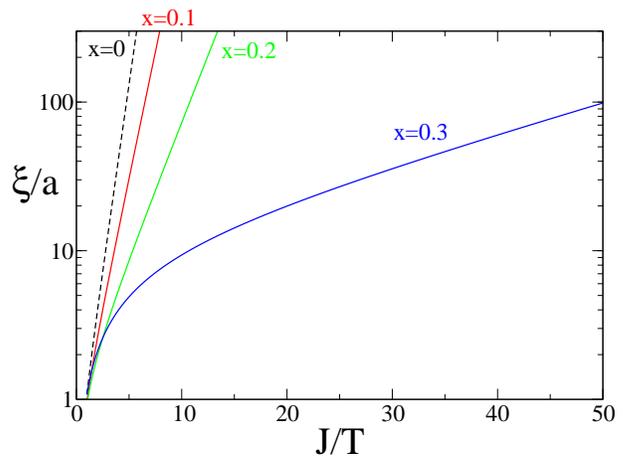}}}
\end{picture}
\caption{$\xi(x,T)$ from Eq. (\ref{xi2}) 
with $\rho_s(x)=\rho_s(0)(1-A_{1/2}\, x)$,
from Eq. (\ref{TN2}), v.s. $J/T$ for $x=0$ (dashed line),
$x=0.1$, $x=0.2$, and $x=0.3$ (solid lines).}
\label{fig_14}
\end{figure}

Our Fig. \ref{fig_14} shows a semi-log plot of $\xi(x,T)$ given by
formula in Eq. (\ref{xi2}) with $\rho_s(x)=\rho_s(0)(1-A_{1/2}\, x)$,
$A_{1/2}$ is from Eq. (\ref{TN2}), v.s. $J/T$ for $x=0$ (dashed line),
$x=0.1$, $x=0.2$, and $x=0.3$ (solid lines). An important observation
can be made here. At small $x$ \, $2\pi\rho_s$ is of the order of $J$
and at all reasonable temperatures the dominant behavior is
exponential in $J/T$ (straight line in the semi-log scale).
When the spin stiffness becomes small ($\rho_s\ll
J$) there is an additional temperature range $J\gg T \gg \rho_s$ where
the exponential behavior is not seen yet while the prefactor gives a
$\log(J/T)$ behavior of the $\log(\xi)$ clearly seen for $x=0.3$.
The experimentally 
observed deviation from the simple exponential behavior
of the correlation length $\xi(T,x)$ v.s. $1/T$ \cite{Greven} can be
related to this effect.
\noindent
\noindent

At larger doping level close to the percolation threshold one expects a
new length-scale to appear. This length-scale is associated with the
crossover from translational invariance to the self-similarity in the
percolative systems \cite{orbach}. Below $x_p$ the ``geometrical length'',
$\xi_G\propto |x-x_p|^{-\nu}$, separates the regions of Euclidean and
fractal geometry. Above $x_p$, where no infinite clusters left, $\xi_G$ is
the characteristic size of the finite clusters. Earlier experimental
studies of the 2D and 3D Ising (Rb$_2$Co$_x$Mg$_{1-x}$F$_4$ and
Fe$_{1-x}$Zn$_x$F$_2$) \cite{Birgeneau1,Birgeneau2} and near-Heisenberg
(Rb$_2$Mn$_x$Mg$_{1-x}$F$_4$ and Mn$_{1-x}$Zn$_x$F$_2$) 
\cite{Birgeneau3,Birgeneau4} systems close to $x_p$ have demonstrated that
the static structure factor ${\cal S}({\bf q})$ contains contributions
from both ``thermal'' and ``geometrical'' lengths in agreement with the
theoretical studies \cite{stinchcombe,Polatsek}. The experimental data
suggest that these lengths combine in the simplest possible form
$\xi^{-1}=\xi^{-1}(T)+\xi^{-1}_G(x)$ and that the Lorentzian form of the
structure factor near ordering vector is preserved. Yet another
interesting result of the proximity to the percolation is that at $x<x_p$
below $T_N$ the elastic Bragg peak should be accompanied by the Lorentzian
whose width at $T=0$ is solely defined by the inverse "geometrical length"
$\xi^{-1}_G$. One would expect similar effects to be observed in newly
available LCO systems close to $x_p$ \cite{Greven}.

It is not clear, however, whether the localization effects in the
infinite cluster, which we discuss in this work, can manifest
themselves in the static structure factor or correlation length. Such
contributions, if they exist, may lead to a new behavior of
correlation length, different from the simple renormalization of
spin stiffness. However, in our approach the potential sources of such
anomalous terms appear in the higher order in $x$ ($\sim x^2$) and,
most certainly, do not affect the results for the experimentally
reachable domain of lengths $\xi\alt 200 a$ above which the ordering
occurs. 
At larger concentrations $x$ such contributions can become important
for shorter correlation lengths 
but in reality they might be screened by the similar effects
from the decoupled clusters.

Theoretically, it is very intriguing if such localization effects of
the infinite cluster can really affect the behavior of correlation
length. We reserve this subject for the further study. 

\section{Conclusions}
\label{conclusions} 

We have studied the problem of diluted 2D and quasi-2D
quantum Heisenberg antiferromagnets in a tetragonal lattice
making use of linear spin-wave theory and  $T$-matrix
approach. We have shown, contrary to the earlier findings,  that the
2D is {\it not} the lower critical dimension for this kind
of disorder and that at $T=0$ long-range order persists up to
concentrations 
close to the classical percolation threshold. These results are
consistent with Monte Carlo simulations in large lattices \cite{Kato}. 
In agreement
with earlier works on this subject, which
studied the problem in the leading order of the 
dilution fraction $x$ \cite{Harris,Wan}, we found that the spin-wave 
spectrum is strongly modified by
disorder.  However, 
contrary to these works we have shown that this result does
{\it not} imply an instability of the system to a paramagnetic phase.
It rather indicates magnon localization on a length scale $\ell$,
exponentially large in $1/x$. We have shown that this new length-scale
appears explicitly in the dynamic properties such as the dynamical
structure factor ${\cal S}({\bf k},\omega)$, Eqs. 
(\ref{A_le_1})-(\ref{S_le_2}),
which can be measured directly in
neutron scattering experiments, and the magnon density of states
$N(\omega)$,
Eqs. (\ref{N_w2}), (\ref{N_w3}) which 
is directly related to the magnetic specific heat Eqs. (\ref{dC_T}),
(\ref{dC_T1}). The measurement
of such quantities will provide a direct test of our theory. Furthermore,
we show that the static properties such as the zero-temperature
staggered magnetization $M(x)$, Eq. (\ref{M}) and N\'eel temperature 
$T_N(x)$, (in the quasi-2D case) do not show any anomaly associated 
with the spectrum and are finite up to the concentration close to the 
classical percolation threshold. These results are in a quantitative
agreement with the 
NQR \cite{Corti}, $\mu$SR \cite{Carretta}, 
ESR \cite{Clarke}, and magnetic susceptibility \cite{Hucker}
measurements in different compounds as well as with the Monte Carlo
data \cite{Kato}. 

We have shown that the effect of dilution of an AF with non-magnetic
impurities is quite strong because dilution removes 
completely spin degrees of freedom from the impurity site
and, therefore, the spin waves
are strongly scattered. Moreover, the low dimensionality of the system 
constraints significantly the phase space for scattering leading to
the localization effects. We have shown that the
hydrodynamic description of the problem breaks down for length-scales
larger than $\ell$ and the spin excitations become diffusive instead of
ballistic. The conventional
averaging procedure which is used to treat disorder  
does not lead to an effective medium with renormalized parameters.
Therefore, one needs to use a different approach for length-scales
larger than $\ell$, the problem which is beyond the scope of this paper.  

In fact, the physics
of localization described in our work 
has similarities to the Anderson localization
for non-interacting electrons in disordered lattices where the
statistics of the excitations does not matter \cite{Anderson}. 
Note that our problem
should be close to the problem of localization of relativistic 
bosons (with chemical potential $\mu=0$) in a random potential. 
On the other hand, that problem is related to the 
problem of disorder in Bose-Hubbard model where non-relativistic 
bosons with kinetic energy $J$ interact through the 
local Coulomb term $U$ \cite{fisher}. 
In the latter model the Bose glass phase appears for small $J$ at 
zero chemical potential, and transition into a superfluid state
is possible when $J$ is large enough. In our case superfluidity is not
possible but we may conjecture that our localized phase is somewhat
similar to the Bose glass phase and magnons are trapped
in the regions which are more ordered than in average. 
It is not clear, however, if the relativistic nature of the bosons 
is important for the nature of localization.

Furthermore, we find the close similarity of our problem 
to the problem of disorder in 2D $d$-wave superconductors
\cite{lee,Yashenkin}. The large enhancement of the density of states
at low frequencies in our case, which comes about because of the
redistribution of spectral weight over the entire Brillouin zone, 
is reminiscent of that problem. 
In $d$-wave superconductors
the elementary excitations are nodal quasiparticles, or relativistic 
(Dirac) fermions. It is known that for these 
excitations localization occurs on a length scale $\ell_L$ 
(localization length)
which is an exponential function of the conductance $\sigma$: $\ell_L
\propto e^{\sigma/\sigma_0}$ \cite{lee_dirac} where $\sigma_0 = e^2/h$. 
Since in the dilute limit one expects the
conductance to diverge with $x$ (that is, $\sigma \propto 1/x$) 
the localization length has the same type of non-analytic dependence on 
$x$ as in our case. However, it is not clear how (if possible) the two 
problems map onto each other. A further investigation, beyond the scope
of this paper, can clarify the connection of the diluted antiferromagnet
with other similar problems of disorder in low-dimensional systems.

In summary, we have presented a comprehensive study of 
diluted quantum Heisenberg
antiferromagnets in 2D and quasi-2D. We have shown that while the dynamic
properties possess anomalies associated with magnon localization the static
properties are free from such anomalies. Thus, in low-dimensional 
systems with disorder
the connection between static and dynamic quantities is not
straightforward. We have compared our results to the numerical simulations
and experimental data with a very
good agreement. We have also proposed other
experiments which can further test the results of our theory. 
Altogether this provides a self-consistent picture of the effects of
disorder in low-dimensional quantum antiferromagnets.

\acknowledgments

We would like to acknowledge invaluable conversations with
A.~Abanov,
A.~Balatsky,
A.~Bishop, 
P.~Carretta,
C.~Chamon,
A.~Chubukov,
R.~Fishman,
B.~Harris,
A.~Harrison,
I.~Gruzberg,
M.~Greven,
M.~Katsnelson,
B.~Keimer,
H.~Mook,
D.~Mandrus,
S.~Nagler,
N.~Nagaosa,
L.~Pryadko,
S.~Sachdev,
A.~Sandvik,
O.~Starykh,
O.~Sushkov,
K.~Takeda,
S.~Todo,
O.~Vajk,
G.~Vignale, 
and M.~Zhitomirsky.
This research was supported in part by 
Oak Ridge National Laboratory, 
managed by UT-Battelle, LLC, for the U.S. Department of Energy under
contract DE-AC05-00OR22725, and by a
CULAR research grant under the auspices of the US Department of Energy.
 
\appendix 
\section{Tetragonal lattice group theory} 
\label{app_A} 

We first resolve the scattering 
potential $\hat{V}$ (\ref{8}), (\ref{9}) 
in \textbf{r}-space by inserting the closure relations \cite{Tinkham}:
\begin{eqnarray}
\hat{V}_{{\bf k}_1,{\bf k}_2} &=&\int d{\bf r}_1\int d{\bf
r}_2\phi_{{\bf k}_1}^{\ast}({\bf r}_1)\hat{V}_{{\bf r}_1,
{\bf r}_2}\phi_{{\bf k}_2}({\bf r}_2)  \label{a1} \\
&=&\int d{\bf r}_1\int d{\bf r}_2\phi_{{\bf k}_1}^{\ast}(
{\bf r}_1)U_i^{\dag}\hat{V}_{{\bf r}_1,{\bf r}_2}
U_i\phi_{{\bf k}_2}({\bf r}_2),  \nonumber
\end{eqnarray}
where $U_{i}$ is any symmetric operator in the group of tetragonal symmetry,
and $\phi_{\bf k}\left({\bf r}\right)$ is a plane wave function, 
$\phi_{\bf k}\left({\bf r}\right) =\left(2\pi \right) ^{3/2}e^{i
\mathbf{k\cdot r}}$, which can be decomposed by projection operators:
\begin{equation}
\phi_{\bf k}({\bf r})=\sum\limits_p\sum\limits_{n=1}^{l_p}\phi
_{\bf k}({\bf r})_n^{(p)},  \label{a2}
\end{equation}
where
\begin{equation}
\phi_{\bf k}({\bf r})_n^{(p)}=\frac{l_p}{g}\sum
\limits_{i=1}^{16}D^{(p)}(U_i)_{nn}U_i\phi_{\bf k}({\bf r}),
\label{a3}
\end{equation}%
where the set of functions, $\{\phi_{\bf k}({\bf r})_n
^{(p)}\}_{n=1,...,l_{p}}$, form a basis of the p$^{th}$ irreducible
representation, and $l_{p}$ is the dimension of p$^{th}$ irreducible
representation; $D^{(p)}(U_{i})_{nn}$ is the diagonal matrix elements of the
p$^{th}$ irreducible representation for the symmetric operator $U_{i}$ in
point group $D_{4h}$ whose order is $g$ ($=16$). We readily project the
potential into irreducible representations as $\hat{V}_{{\bf k}_1,
{\bf k}_2}=\sum_p\hat{V}_{{\bf k}_1,{\bf k}_2}^{(p)}$, where
\begin{eqnarray}
\hat{V}_{{\bf k}_1,{\bf k}_2}^{(p)}
&=&\sum_{n=1}^{l_{p}}\sum_{i,j=1}^g\frac{l_{p}^{2}}{g^{2}}
D^{(p)}(U_i)_{nn}D^{(p)}(U_j)_{nn}  \label{a4} \\
&&\times \int d{\bf k}_3\int d{\bf k}_4 A_{{\bf k}_1, {\bf k}_3}^i
\hat{V}_{{\bf k}_3,{\bf k}_4}A_{{\bf k}_4,{\bf k}_2}^j,  \nonumber
\end{eqnarray}
where
\begin{equation}
A_{{\bf k}_1,{\bf k}_2}^i=\int d{\bf r}\phi_{{\bf k}_1}^{\ast}
({\bf r})U_i\phi_{{\bf k}_2}({\bf r}).  \label{a5}
\end{equation}
Using the tetragonal symmetry group 
one notices that each $A_{{\bf k}_1,{\bf k}_2}^i$ 
is a $\delta $-function. Thus, the scattering potential 
$\hat{V}_{{\bf k}_1,{\bf k}_2}^A$ Eq. (\ref{8}) can be decomposed into
channels of irreducible representations. 
The non-zero orthogonal channels are (before the Bogolyubov
transformation):

A$_{1g}$ ($s$-wave): 
\begin{equation}
\hat{V}_{{\bf k}_1,{\bf k}_2}^{A,s}=\left| \mbox{s}_{{\bf k}
_1}\rangle\otimes \langle \mbox{s}_{{\bf k}_2}\right| +\left| \mbox{s}_{
{\bf k}_1}^{\perp}\rangle\otimes\langle \mbox{s}_{{\bf k}
_2}^{\perp }\right|\ ,  \label{a6}
\end{equation}

E$_{u}$ (in-plane $p$-waves):
\begin{equation}
\hat{V}_{{\bf k}_1,{\bf k}_2}^{A,p_{x(y)}}=\left| \mbox{p}_{{\bf k}
_1}^{x(y)}\rangle \otimes \langle \mbox{p}_{{\bf k}_2}^{x(y)}\right|
\ ,  \label{a7}
\end{equation}

B$_{1g}$ ($d$-wave):
\begin{equation}
\hat{V}_{{\bf k}_1,{\bf k}_2}^{A,d}=\left| \mbox{d}_{{\bf k}
_1}\rangle\otimes\langle\mbox{d}_{{\bf k}_2}\right| ,
\label{a8}
\end{equation}

A$_{2u}$ ($p_z$-wave):
\begin{equation}
\hat{V}_{{\bf k}_1,{\bf k}_2}^{A,p_z}=\left| \mbox{p}_{{\bf k}_1}^z
\rangle \otimes \langle \mbox{p}_{{\bf k}_2}^z\right| ,
\label{a9}
\end{equation}
where $\langle \mbox{s}_{\bf k}| =[1,\, \gamma_{\bf k}]$, $
\langle \mbox{s}_{\bf k}^\perp| =\sqrt{\tau}\, [1,\, \gamma_{\bf
k}^\perp]$, $\langle \mbox{p}_{\bf k}^{x(y)}| = [0,\, 1]\, \sin
k_{x(y)}/\sqrt{2}$, $\langle \mbox{d}_{\bf k}| =[0,\, 1]\, \gamma_{
\bf k}^-$ and $\langle \mbox{p}_{\bf k}^z| =\sqrt{\tau}\,
\sin k_z\, [0,\, 1]/\sqrt{2}$. Bogoliubov transformation yields 
Eqns. (\ref{V_s})-(\ref{V_pz}).

 
\section{3D T-matrix} 
\label{app_B} 
In this Appendix we provide the solution of the $s$-wave $T$-matrix 
equation   
in the tetragonal lattice for the arbitrary relative value of the  
inter-plane and in-plane exchange integrals $\tau=J_\perp/2J$. 
With this solution we demonstrate the smallness of the 
3D corrections to the 2D result in the quasi-2D case ($\tau\ll 1$). 
 
After some algebra one can solve the $T$-matrix equation (\ref{T}) with the  
$s$-wave scattering potential from Eq. (\ref{V_s}) (sublattice $A$),  
$u_{\bf k}$, $v_{\bf k}$, and $\omega_{\bf k}$ from Eqs. (\ref{BT1}), 
(\ref{wk}) and obtain: 
\begin{eqnarray} 
\label{T_3D} 
&&\hat{T}^{A,s}_{{\bf k},{\bf k}^\prime}(\omega)  
=|s_{\bf k}\rangle\otimes\langle s_{{\bf k}^\prime}|\, \cdot 
\, \Gamma_1(\omega) 
\nonumber\\ 
&&\phantom{\hat{T}^{A,s}_{{\bf k},{\bf k}^\prime}(\omega)=} 
+ |s^\perp_{\bf k}\rangle\otimes\langle s^\perp_{{\bf k}^\prime}| 
\, \cdot\, \Gamma_2(\omega)\\ 
&&\phantom{\hat{T}^{A,s}_{{\bf k},{\bf k}^\prime}(\omega)=} 
+\left( |s_{\bf k}\rangle\otimes\langle s^\perp_{{\bf 
k}^\prime}|+ |s^\perp_{\bf k}\rangle\otimes\langle s_{{\bf k}^\prime}|\right) 
\, \cdot\, \Gamma_3(\omega) \ ,\nonumber 
\end{eqnarray} 
where the $\omega$-dependence of the in-plane scattering (first term) 
is given by expression which is formally similar to the pure 2D result 
Eq. (\ref{gammas}): 
\begin{eqnarray} 
\label{gamma_1} 
\Gamma_1(\omega)=\frac{1}{\omega}+\frac{(1+\omega)\rho(\omega)+\tau R_1(\omega)} 
{1-\omega(1+\omega)\rho(\omega)+\tau D(\omega)}\ , 
\end{eqnarray} 
with $\rho(\omega)$ given by Eq. (\ref{rhos}). Note that the 
integration over ${\bf p}$ in this case is three-dimensional. 
The inter-plane scattering is: 
\begin{eqnarray} 
\label{gamma_2} 
\Gamma_2(\omega)= -\tau+\frac{\tau^2}{\omega}-\frac{\tau^2 R_2(\omega)} 
{1-\omega(1+\omega)\rho(\omega)+\tau D(\omega)}\ . 
\end{eqnarray} 
The $\omega$-dependence of the cross-term is given by: 
\begin{eqnarray} 
\label{gamma_3} 
\Gamma_3(\omega)=\frac{\tau}{\omega}+\frac{\tau R_3(\omega)} 
{1-\omega(1+\omega)\rho(\omega)+\tau D(\omega)}\ . 
\end{eqnarray} 
All three parts of the scattering matrix   
possess the same ``unphysical'' $1/\omega$ contribution discussed in 
the text. Application of the projection procedure Eqs. (\ref{D_H}), 
(\ref{D_V_s}) to this problem is out of the scope of this Appendix. 
 
The auxiliary functions $D$ and  $R_i$ are given by rather cumbersome  
combinations of $\omega$, $\rho(\omega)$ and two additional integrals: 
\begin{eqnarray} 
\label{alp_bet} 
\alpha(\omega)=\sum_{\bf p}\frac{(\gamma^\perp_{\bf p})^2} 
{\omega^2-\omega_{\bf p}^2}\ , \ \ 
\beta(\omega) 
=\sum_{\bf p}\frac{\hat{\gamma}_{\bf p}\gamma^\perp_{\bf p}} 
{\omega^2-\omega_{\bf p}^2}\ , 
\end{eqnarray} 
with $\hat{\gamma}_{\bf p}$, $\gamma^\perp_{\bf p}$, $\omega_{\bf p}$ 
from Eqs. (\ref{hat_gamma}), (\ref{wk}). Note that at $\tau\rightarrow 
0$ $\alpha(\omega)\rightarrow \rho(\omega)/2$ and  
$\beta(\omega)\rightarrow 0$. 
 
The expressions for $D$ and  $R_i$ are: 
\begin{eqnarray} 
\label{aux} 
&&D=P-\omega P_2 \ , \nonumber\\ 
&&R_1=\rho-\alpha+P_2\ , \\ 
&&R_2=\alpha-(\omega-\tau)P_2\ , \nonumber\\ 
&&R_3=(\rho+\alpha-P)/2+\tau(\rho-\alpha)/2+P_2 
\ , \nonumber 
\end{eqnarray} 
where the following shorthand notations are used 
\begin{eqnarray} 
\label{Ps} 
&&P=\hat{\gamma}_0 (\rho +\alpha)-2\beta \ , \\ 
&&P_2=\hat{\gamma}_0^2\rho\alpha-\beta^2+\alpha-\omega P
-\omega^2\rho\alpha\ . \nonumber 
\end{eqnarray} 
There is no assumption on the value of $\tau$ made in these formulae. 
 
At $\tau\ll 1$ and $\omega\ll 1$ ($\omega$ can be still $\gg \tau$) 
one can show that: 
\begin{eqnarray} 
\label{aux1} 
&&D\simeq 3\rho/2 \ , \ \ R_1\simeq\rho^2/2+\rho\ ,\nonumber \\ 
&&R_2\simeq \rho/2 \ , \ \ R_3={\cal O}(\tau\rho^2)\ .  
\end{eqnarray} 
In the same limit $\tau\ll 1$ and $\omega\ll 1$  
the $\omega$-dependent parts of the scattering 
matrix become (we simply omit the unphysical $1/\omega$ terms here): 
\begin{eqnarray} 
\label{gamma_limit} 
&&\Gamma_1(\omega)\simeq \rho-\tau (\rho^2-\rho)\ 
,\nonumber \\ 
&&\Gamma_2(\omega)\simeq -\tau+\tau^2\rho/2\
, \\ 
&&\Gamma_3(\omega)= {\cal O}(\tau^2\rho^2)\ .\nonumber  
\end{eqnarray} 
Recall that in 2D $\Gamma_1(\omega)\simeq \rho$ and
$\Gamma_2(\omega)=\Gamma_3(\omega)\equiv 0$.
Since Re$\rho\sim \ln|\omega|$ at $\omega \gg \sqrt{\tau}$ 
and Re$\rho\sim \ln|\tau|$ at $\omega \le \sqrt{4\tau}$ the 
largest relative correction to the 2D terms in the scattering matrix  
is ${\cal O}(\tau\ln(\tau))$. The same statement can be proved for all 
higher powers of $\omega$ in Eq. (\ref{gamma_1}) without making 
$\omega\ll 1$ assumption.  
 
The conclusion is, once again, that at 
$\tau\ll 1$ one can safely drop all terms explicitly proportional to  
$\tau$ in Eqs. (\ref{gamma_1})-(\ref{gamma_3})  and thus arrive to the 
purely 2D expression for the scattering matrix given in
Eq. (\ref{gammas}). The only modification in the quasi-2D case 
versus 2D case is the change of the behavior of  $\rho(\omega)$ at low 
$\omega$, whose real part saturates at $\omega\le \sqrt{4\tau}$ and 
imaginary part acquires an extra power in $\omega$ (see Appendix
\ref{app_E}).  

\section{Elliptic integrals} 
\label{app_C} 

The energy-dependent part of the $T$-matrix Eqns. (\ref{gammas}), 
(\ref{gammas}) is expressed through the integrals of the Green's
functions Eqn. (\ref{rhos}). These integrals can be evaluated in the
case of 2D and are given by combinations of complete elliptic
integrals of the first and second kind:
\begin{eqnarray} 
\label{rhos_1} 
&&\rho(\omega)=\sum_{\bf p}\frac{1}{\omega^2-\omega_{\bf p}^2}
=-\frac{2}{\pi\omega^\prime}\bigg[K(\omega^\prime)+iK(\omega)
\bigg]
\ , \ \ \\
&&\rho_d(\omega) 
=\sum_{\bf p}\frac{(\gamma^-_{\bf p})^2}{\omega^2-\omega_{\bf p}^2}
=1+  \frac{2}{\pi\omega^\prime} \bigg[\omega^2
K(\omega^\prime)-2E(\omega^\prime)\nonumber\\ 
&&\phantom{\rho(\omega)=1+  \frac{2}{\pi\omega^\prime} \bigg[}
+i\bigg((\omega^2-2)K(\omega)+E(\omega)\bigg)\bigg] \nonumber ,
\end{eqnarray} 
where $\omega^\prime=\sqrt{1-\omega^2}$, $K$ and $E$ are the  complete
elliptic integrals of the first and second kind, respectively \cite{GR}.

In the low-energy limit:
\begin{eqnarray} 
\label{rhos_2} 
&&\rho(\omega)=\frac{2}{\pi}\ln|\omega/4|-i \ ,  \\
&&\rho_d(\omega)=1- \frac{4}{\pi} \nonumber \ .
\end{eqnarray} 
 
\section{Projection of unphysical states} 
\label{app_D} 

After introduction of the fictitious magnetic field to
project out the unphysical on-site mode the $s$-wave scattering
potential (sublattice $A$) is given by the sum of two terms from
Eqs. (\ref{V_sa}), (\ref{D_V_s}):
\begin{eqnarray} 
\label{V_s_tot} 
&&{\hat{\cal V}}^{A,s,total}_{{\bf k},{\bf k}^\prime}=-|s_{\bf k}\rangle 
\otimes\langle s_{{\bf k}^\prime}|+H_z|\Delta s_{\bf k}\rangle 
\otimes\langle \Delta s_{{\bf k}^\prime}|\ ,\nonumber\\ 
&& \ \ 
\mbox{with} \ \ \langle s_{\bf k}|=\omega_{\bf k} 
\big[u_{\bf k},\ \ -v_{\bf k}\big]\ , \ \   \langle \Delta s_{\bf k}|=  
\big[u_{\bf k},\ \ v_{\bf k}\big] \ . 
\end{eqnarray} 
One immediately suggest the form of the solution of the $T$-matrix
equation: 
\begin{eqnarray} 
\label{T_s_tot} 
&&\hat{T}^{A,s}_{{\bf k},{\bf k}^\prime}(\omega)  
=|s_{\bf k}\rangle\otimes\langle s_{{\bf k}^\prime}|\, \cdot 
\, \Gamma_1(\omega) \nonumber\\
&&\phantom{\hat{T}^{A,s}_{{\bf k},{\bf k}^\prime}(\omega)=} 
+|\Delta s_{\bf k}\rangle\otimes\langle \Delta s_{{\bf k}^\prime}| 
\, \cdot\, \Gamma_2(\omega) \\ 
&&\phantom{\hat{T}^{A,s}_{{\bf k},{\bf k}^\prime}(\omega)=} 
+\left(  | \Delta s_{\bf k}\rangle\otimes\langle s_{{\bf 
k}^\prime}|+ |s_{\bf k}\rangle\otimes\langle \Delta 
s_{{\bf k}^\prime}|\right) \, \cdot\, \Gamma_3(\omega) \ ,\nonumber 
\end{eqnarray} 
and, after some algebra, one finds:
\begin{eqnarray} 
\label{gamma_H} 
&&\Gamma_1(\omega)=\frac{H_z (1+\omega)\rho(\omega)+1} 
{[1-\omega(1+\omega)\rho(\omega)](H_z-\omega)}\ ,\nonumber\\ 
&&\Gamma_2(\omega)= -\frac{\omega H_z}{H_z-\omega}\ ,\\
&&\Gamma_3(\omega)=\frac{H_z}{H_z-\omega}\ ,\nonumber
\end{eqnarray} 
which yield the answer given in Eqs. (\ref{T_2})-(\ref{D_T}) in the
limit $H_z\rightarrow\infty$.
 
\section{3D $\rho(\omega)$} 
\label{app_E} 

The key ingredient of the low-energy 
$T$-matrix scattering is given by the integral of the Green's 
function over ${\bf k}$, $\rho(\omega)$ (\ref{rhos}). 
Appendix \ref{app_C} gives an analytical expression of $\rho(\omega)$ 
in the 2D case. In the quasi-2D case the inter-plane coupling provides
a cut-off in the logarithm and gives an extra power of $\omega$ in the
imaginary part of the integral in the 3D energy range. 
This can be obtained explicitly using 3D form of the spin-wave
dispersion Eq. (\ref{wk}) $\omega _{{\bf
k}}=\sqrt{\hat{\gamma}_0^{2}-\hat{\gamma}_{\bf k}^{2}}$. 

In the limit $\sqrt{\tau}=\sqrt{J_\perp/2J} \ll 1$, and $\omega\ll 1$
(for arbitrary $\omega/\sqrt{\tau}$) one obtains for the real part of 
$\rho(\omega)$:
\begin{eqnarray} 
\label{rho_3D_1} 
&&\mbox{Re}\rho(\omega)= \frac{2}{\pi}\ln\bigg|
\frac{\sqrt{\tau}}{4}\bigg| +{\cal
O}(\tau,\omega^2),  \ \ \mbox{for}\ \ \omega\le \sqrt{4\tau}\ ,\nonumber \\
&&\mbox{Re}\rho(\omega)= \frac{2}{\pi}\ln\bigg|
\frac{\omega+\sqrt{\omega^2-4\tau}}{8}\bigg| +{\cal
O}(\tau,\omega^2), \\
&&\phantom{\mbox{Re}\rho(\omega)= \frac{2}{\pi}\ln\bigg|
\frac{\omega+\sqrt{\omega^2-4\tau}}{8}\bigg|} 
\ \ \mbox{for}\ \ \omega\ge \sqrt{4\tau}\ , \nonumber 
\end{eqnarray} 
at $\omega\gg \sqrt{4\tau}$ the 3D energy scale is irrelevant and 
Re$\rho(\omega)=\frac{2}{\pi}\ln|\omega/4|$  is back to its 2D form.
Imaginary part of $\rho(\omega)$ is
\begin{eqnarray} 
\label{rho_3D_2} 
&&\mbox{Im}\rho(\omega)= -\frac{1}{\pi}\arccos\bigg(
\frac{\sqrt{(1+\tau)^2-\omega^2}-1}{\tau}\bigg)
+{\cal O}(\omega^2)\nonumber\\
&&\phantom{\mbox{Im}\rho(\omega)}
= -\frac{1}{\pi}\arccos\bigg(
1-\frac{\omega^2}{2\tau}\bigg)+{\cal O}(\tau^2,\omega^2)
, \nonumber\\ 
&&\phantom{\mbox{Im}\rho(\omega)-\frac{1}{\pi}\arccos\bigg(
1-\frac{\omega^2}{2\tau}\bigg)}
\ \ \mbox{for}\ \ \omega\le \sqrt{4\tau}\ ,\\
&&\mbox{Im}\rho(\omega)= -1+{\cal O}(\omega^2),  \ \ \mbox{for}\ \ 
\omega\ge \sqrt{4\tau}\ , \nonumber 
\end{eqnarray} 
at small $\omega\ll \sqrt{4\tau}$ deep into the 3D range of energies
Im$\rho(\omega)=-\frac{\omega}{\pi\sqrt{\tau}}$ is linear in
$\omega$.  

\section{$T_N(_x)$ for the Ising problem}
\label{app_F} 

In this Appendix we apply the formalism of our work to the problem of
$T_N(x)$ v.s. $x$ dependence for the Ising $S=1/2$ case. While the
spin-wave approximation is much less adequate in the Ising limit 
than for the pure Heisenberg model it is nevertheless 
a very instructive exercise
which gives a quantitatively correct answer. 

Quadratic part of the 2D $S=1/2$ Ising model in the spin-wave
approximation reads as
\begin{eqnarray} 
\label{HI} 
&&\frac{{\cal H}}{2J}={\cal H}_0+{\cal H}_{imp} \\
&&\phantom{\frac{{\cal H}}{2J}}
=\sum_{\bf k} a^{\dag}_{\bf k}a_{\bf k}-
\sum_{l,{\bf k},{\bf k}^\prime} 
e^{i({\bf k}-{\bf k}^\prime){\bf R}_l} 
V_{{\bf k},{\bf k}^\prime}
a_{\bf k}^{\dag} a_{{\bf k}^\prime}\ , \nonumber
\end{eqnarray} 
with
\begin{eqnarray} 
\label{VI} 
V_{{\bf k},{\bf k}^\prime}=\gamma_{{\bf k}-{\bf k}^\prime} \ ,
\end{eqnarray} 
where we omit from the beginning the ``unphysical'' term which will
result in $\omega=0$ mode.
$T$-matrix gives the total result for all scattering channels:
\begin{eqnarray} 
\label{TI} 
T^{tot}_{{\bf k},{\bf k}^\prime}(\omega)=-\gamma_{{\bf k}-{\bf
k}^\prime}\frac{\omega-1}{\omega-3/4}\ ,
\end{eqnarray} 
where we used the property
\begin{eqnarray} 
\label{addI} 
\sum_{\bf p}\gamma_{{\bf k}-{\bf p}}\gamma_{{\bf p}-{\bf k}^\prime}
\equiv \gamma_{{\bf k}-{\bf k}^\prime}/4 \ .
\end{eqnarray} 
The self-energy is then given by:
\begin{eqnarray} 
\label{SI} 
\Sigma(\omega)=-x\,\frac{\omega-1}{\omega-3/4}\ .
\end{eqnarray} 
The Green's function has two poles now:
\begin{eqnarray} 
\label{GI} 
G(\omega)=\frac{1}{\omega-1}\cdot\frac{\omega-3/4}{\omega-3/4+x}\ ,
\end{eqnarray} 
and the spectral function is given by two $\delta$-peaks:
\begin{eqnarray} 
\label{AI} 
A(\omega)=\frac{1}{1+4x}\big[\delta(\omega-1)+
4x\delta(\omega-3/4+x)\big]\ .
\end{eqnarray} 

N\'eel temperature is defined from the condition:
\begin{eqnarray} 
\label{TNI1} 
\langle S^z\rangle (T_N,x)=\frac{1}{2}-
\int_{-\infty}^{\infty}d\omega\  
n_B(\omega)A(\omega)=0 \ ,
\end{eqnarray} 
which transforms to 
\begin{eqnarray} 
\label{TNI2} 
\frac{1+4x}{2}=n_B(1)+4x\, n_B(3/4-x) \ .
\end{eqnarray} 
In a pure system 
$T_N(0)/2J=1/\ln 3$. At small $x$ \ $T_N(x)\simeq T_N(0) (1-A^I
x)$ and, after some algebra, one obtains an analytical expression for
$A^I$ 
\begin{eqnarray} 
\label{A_I} 
A^I=\frac{4}{3}\, \frac{2}{\ln3}\left[\frac{2}{3^{3/4}-1}
-1\right]\simeq 1.025 \, \frac{4}{3}\simeq 1.37\ ,
\end{eqnarray} 
which should be compared with the RPA answer $A^I_{RPA}=4/3$ \cite{McGurn}
and an exact answer $A^I_{exact}\simeq 1.57$  \cite{McGurn1}.
One can see that in spite of the roughness of the approximation of the
Ising spin degrees of freedom by bosons our approach
gives a good quantitative agreement with other approaches and an 
exact result.
 


\begin{references}
\bibitem{remark} A similar statement has been made in Ref. \cite{Wan}.
\bibitem{remark1} We use $M(x)=\sum_i |S^z_i|/N_m$, where $N_m$ is the  
number of magnetic sites. Another physically meaningful normalization  
is by the number of magnetic sites inside the infinite clusters  
$N_m^{inf}$. These definitions are connected  $M(x)=M_{inf}(x) P(x)/(1-x)$  
and at $x$ not too close to $x_p$ they are equivalent $M(x)\approx  
M_{inf}(x)$ since the probability to find a magnetic site in the  
infinite cluster $P(x)\simeq 1-x-\pi x^2$.  
Sometimes magnetization is normalized by the  
total number of sites $N_{tot}$, like in MC study  
\cite{Kato}. Evidently, $M_{tot}(x)=M(x)\, (1-x)$.
\end{references}
\end{document}